\documentclass[nofootinbib,a4paper,12pt]{article}
\usepackage{jheppub}
\usepackage[T1]{fontenc}
\usepackage[utf8]{inputenc}
\usepackage{float}
\usepackage{slashed}
\usepackage{multicol,multirow}
\usepackage{amssymb}
\usepackage{amsmath}
\usepackage{subcaption}
\usepackage{array}
\usepackage{color}
\usepackage{hyperref}
\hypersetup{colorlinks=true}
\usepackage{longtable}
\usepackage{graphics,graphicx,epsfig,ulem,booktabs}
\numberwithin{equation}{section}

\definecolor{green}{rgb}{0.1,0.8,0.2}
\definecolor{orange}{rgb}{1.0,0.5,0.0}
\definecolor{cyan}{rgb}{0.0,0.75,0.8}

\newcolumntype{C}[1]{>{\centering\let\newline\\\arraybackslash\hspace{0pt}}m{#1}}

\newcommand{\Br}{{\rm Br}}
\newcommand{\Gu}{\Gamma (B^+)}
\newcommand{\Gd}{\Gamma (B_d)}
\newcommand{\Gs}{\Gamma (B_s)}
\newcommand{\tud}{\tau (B^+)/\tau (B_d)}
\newcommand{\tsd}{\tau (B_s)/\tau (B_d)}
\newcommand{\GeV}{\, {\rm GeV}}
\newcommand{\mupi}{\mu_\pi^2 (B)}
\newcommand{\muG}{\mu_G^2 (B)}
\newcommand{\rhoD}{\rho_D^3 (B)}
\newcommand{\mupis}{\mu_\pi^2 (B_s)}
\newcommand{\muGs}{\mu_G^2 (B_s)}
\newcommand{\rhoDs}{\rho_D^3 (B_s)}

\title{Disintegration of beauty: a precision study}

\preprint{SI-HEP-2022-22}

\author[a]{Alexander Lenz,}
\author[a]{Maria Laura Piscopo,}
\author[a]{Aleksey V. Rusov}

\affiliation[a]{Physik Department, Universit\"{a}t Siegen, Walter-Flex-Str. 3, 57068 Siegen, Germany}

\emailAdd{alexander.lenz@uni-siegen.de}
\emailAdd{maria.piscopo@uni-siegen.de}
\emailAdd{rusov@physik.uni-siegen.de}

\abstract{We update the Standard Model (SM) predictions for $B$-meson lifetimes within the heavy quark expansion (HQE). Including for the first time the contribution of the Darwin operator, SU(3)$_F$ breaking corrections to the matrix element of dimension-six four-quark operators and the so-called eye-contractions, we obtain for the total widths
 $\Gamma (B^+) = (0.58^{+0.11}_{-0.07}) \, \mbox{ps}^{-1}$,
$\Gamma (B_d) = (0.63^{+0.11}_{-0.07}) \, \mbox{ps}^{-1}$,
$\Gamma (B_s) = (0.63^{+0.11}_{-0.07}) \, \mbox{ps}^{-1}$, and for the lifetime ratios
 $\tau (B^+) / \tau (B_d) = 1.086 \pm 0.022$,
$\tau (B_s) / \tau (B_d) = 1.003 \pm 0.006
\, (1.028 \pm 0.011)
$.
The two values for the last observable arise from using two
different sets of input for the non-perturbative
parameters $\mu_\pi^2(B_d)$, $\mu_G^2(B_d)$, and
$\rho_D^3(B_d)$  as well as from 
different estimates of the SU(3)$_F$ breaking in these
parameters. Our results are overall in very good
agreement with the corresponding experimental data,
however, there seems to emerge a tension in $\tsd$
when considering  the second set of input parameters.
Specifically, this observable is extremely sensitive 
to the size of the parameter $\rho_D^3 (B_d)$ 
and of the SU(3)$_F$ breaking effects 
 in $\mu_\pi^2$, $\mu_G^2$ and $\rho_D^3$;
hence, it is of utmost importance to be able to better constrain 
all these parameters. In this respect, an extraction of $\mupis, \muGs, \rhoDs$ from future experimental data on inclusive semileptonic $B_s$-meson decays or from direct non-perturbative calculations,
as well as more insights about the value of $\rhoD$ extracted from fit,
would be very helpful in reducing the corresponding theory uncertainties.
}

\begin{document}

\maketitle
\flushbottom

\section{Introduction}
\begin{figure}
    \centering
    \includegraphics[scale=1.0]{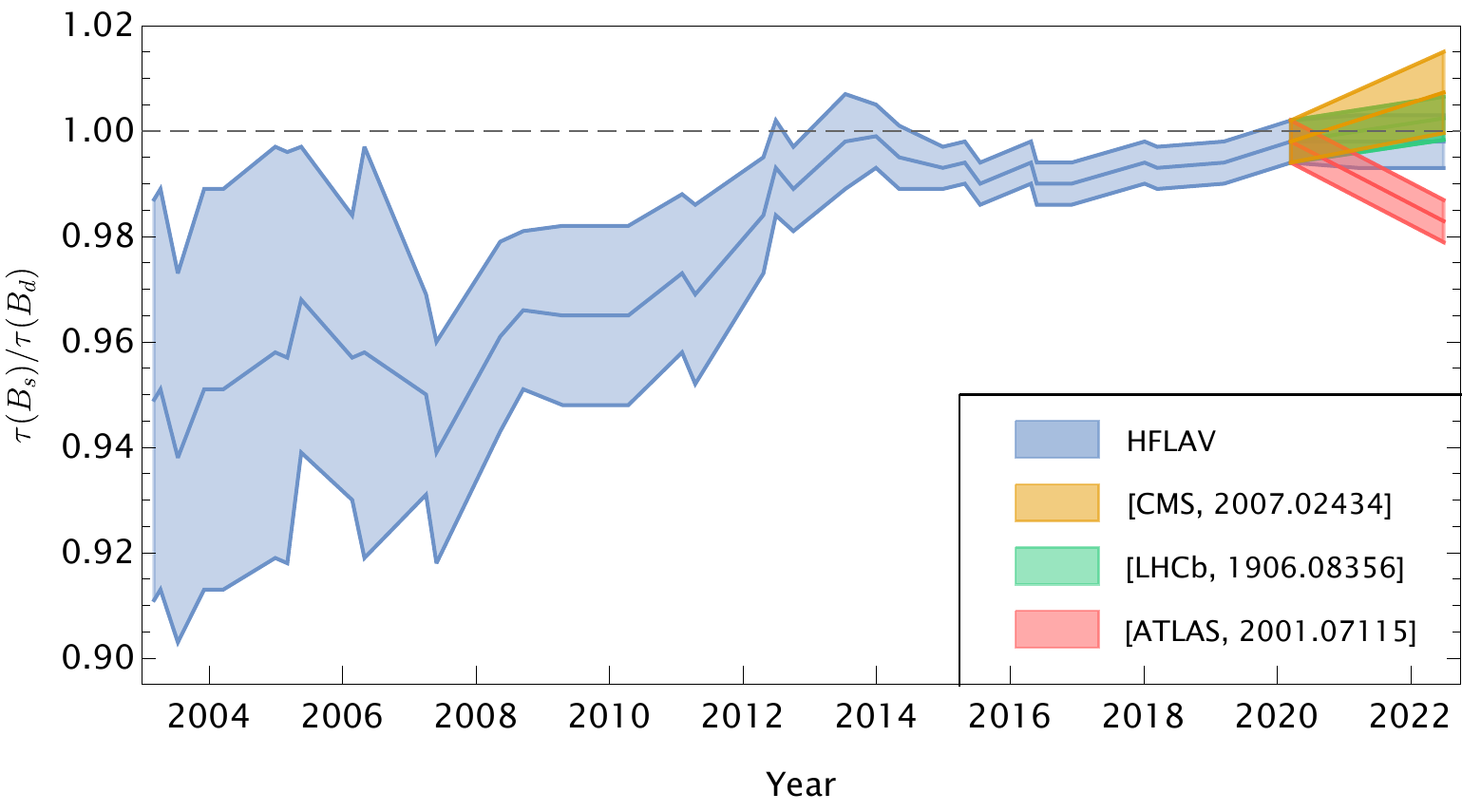} \,
   \caption{ HFAG/HFLAV results for the lifetime ratio $\tau (B_s) / \tau(B_d)$ from 2003 till 2022. Note, that the recent measurements of $\Gamma_s$ by ATLAS (red) seem to deviate from the most recent determinations by LHCb (green) and CMS (orange);  the corresponding bands are in fact obtained by fixing the current HFLAV value for $\tau (B_d)$.
  }
    \label{fig:lifetime_history}
\end{figure}

The total decay width $\Gamma$ or equivalently its inverse, the total lifetime $\tau = \Gamma^{-1}$, defines one of the fundamental properties of elementary and composite particles, and hence represents an observable of phenomenological primary importance. Moreover, in particular for the case of heavy hadrons, the study of lifetimes can provide an interesting playground to test our understanding of the weak and the strong interactions.\\
Experimentally, the lifetimes of  $B$ mesons are determined very precisely by now~\cite{Amhis:2022mac} (based on the measurements in Refs.~\cite{DELPHI:2003hqy,ALEPH:2000kte,ALEPH:1996geg,DELPHI:1995hxy,DELPHI:1995pkz,DELPHI:1996dkh,L3:1998pnf,OPAL:1995bfe,OPAL:1998msi,OPAL:2000qeg,SLD:1997wak,CDF:1998pvs,CDF:2002ixx,CDF:2010ibe,D0:2008nly,D0:2012hfl,D0:2014ycx,BaBar:2001mmd,BaBar:2002nat,BaBar:2002jxa,BaBar:2002war,BaBar:2005laz,Belle:2004hwe,ATLAS:2012cvl,CMS:2017ygm,LHCb:2014qsd,LHCb:2014bqh,CDF:2010gif,D0:2004ije,ALEPH:1997rqk,DELPHI:2000aij,OPAL:1997zgk,CDF:1998htf,DELPHI:2000gjz,OPAL:1997ufs,CDF:2011utg,LHCb:2013cca,LHCb:2014wet,LHCb:2017knt,CDF:1997axv,D0:2004jzq,LHCb:2021awg,CMS:2019bbr,ALEPH:2000cjd,LHCb:2012zwr,LHCb:2013dzm,CDF:2011kjt,D0:2016nbv,LHCb:2016crj,LHCb:2013odx,CDF:2012nqr,D0:2011ymu,ATLAS:2014nmm,ATLAS:2016pno,ATLAS:2020lbz,CMS:2015asi,CMS:2020efq,LHCb:2014iah,LHCb:2017hbp,LHCb:2016tuh}).
\begin{table}[h]
\centering
\renewcommand{\arraystretch}{1.6}
    \begin{tabular}{|c||C{3cm}|C{3cm}|C{3cm}|}
    \hline
         & $B^+$ & $B_d$ & $B_s$ 
         \\
         \hline
         \hline
    $\tau \, [{\rm ps}]$ & $1.638 \pm 0.004 $ & $1.519 \pm 0.004$ & $1.516 \pm 0.006$
    \\
    \hline
     $\Gamma \, [{\rm ps}^{-1}]$ & $0.6105 \pm 0.0015 $ & $0.6583 \pm 0.0017$ & 
                                 $ 0.6596 \pm 0.0026 $
    \\
    \hline
    $\tau (B_q)/\tau (B_d)$ & $1.076 \pm 0.004 $ & $1$ & $ 0.998 \pm 0.005$
    \\
    \hline
    \end{tabular}
    \caption{Status of the experimental determinations of the $B$-meson lifetimes~\cite{Amhis:2022mac}.}
    \label{tab:exp-data}
\end{table}
The values 
in Table \ref{tab:exp-data} show that the lightest $B$~mesons have roughly the same lifetime and that in the ratio
$\tau (B_s) / \tau(B_d)$ all differences seem to almost cancel out. The  improvement of the experimental determination for this ratio 
over the last 20 years can be read off Fig.~\ref{fig:lifetime_history}.
 Interestingly, 
 the recent measurement of $\Gamma_s$ by ATLAS \cite{ATLAS:2020lbz} deviates from the most recent ones by LHCb
 \cite{LHCb:2019sgv,LHCb:2019nin,LHCb:2021wte} and CMS \cite{CMS:2020efq} by $2-4 \sigma$ - an experimental clarification of the origin of these discrepancies
 is of course highly desirable.
\\
On the theoretical side, inclusive decay widths of heavy hadrons can by systematically computed in the framework of the heavy quark expansion (HQE), see e.g. the review \cite{Lenz:2014jha}. 
Predictions for lifetime ratios of $B$
mesons based on this method trace back to the 80s
and a selection of results is given in Table~\ref{tab:SM-results}.
\begin{table}[h]
\centering
\renewcommand{\arraystretch}{1.6}
    \begin{tabular}{|c||C{3cm}|C{3cm}|}
    \hline
         & $\tau (B^+)/\tau (B_d)$  & $\tau (B_s)/\tau (B_d)$ 
         \\
         \hline
         \hline
Shifman, Voloshin, 1986 \cite{Shifman:1986mx} & $\approx 1.1 $ & $\approx 1$
   \\
       \hline
  Neubert, Sachrajda, 1996 \cite{Neubert:1996we} & fixed to 1.02& $1 \pm {\cal O } (1 \%)$
   \\
        \hline
   Gabbiani et al., 2004 \cite{Gabbiani:2004tp} & $1.06 \pm 0.02 $& $1.00 \pm 0.01$ 
   \\
       \hline
   Kirk et al., 2017 \cite{Kirk:2017juj} & $1.082^{+0.022}_{-0.026}$& $1.0007 \pm 0.0025$ 
   \\
       \hline
    \end{tabular}
    \caption{Selected theoretical determinations of the $B$-meson lifetimes.}
    \label{tab:SM-results}
\end{table}
According to the HQE the 
total decay rate of the $B_q$ meson, $\Gamma(B_q)$,
with $q = (u,d,s)$~\footnote{Note that we only consider
bound states of the $b$ and a light quark. In the case of
the $B_c$ meson the HQE must be properly generalised to
describe two weakly decaying heavy quarks,
see Refs.~\cite{Beneke:1996xe, Aebischer:2021ilm, Aebischer:2021eio}.
 }, 
can be expressed as a series expansion
in inverse powers of the heavy quark mass. 
Due to the large value of the $b$-quark mass in
comparison to a typical hadronic scale, we expect
a good convergence of the HQE for $B$ mesons.
Moreover, recent studies of $D$ meson lifetimes
\cite{charm,Gratrex:2022xpm} and charmed baryon lifetimes
\cite{Gratrex:2022xpm} indicate a convergence of the HQE
even in the charm system.
\\
In the framework of the HQE, $\Gamma(B_q)$ can be
split up into the sum of a leading contribution
stemming from the decay
of the free heavy $b$-quark, $\Gamma_b$, and of a subleading one which is specific to the $B_q$ meson, $\delta \Gamma_{B_q} $:
\begin{equation}
  \Gamma (B_q)  =  \Gamma_b + \delta \Gamma_{B_q} \, .
\end{equation}
We stress that the free quark decay is proportional to the factor $\Gamma_0
= G_F^2 m_b^5 |V_{cb}|^2/(192 \pi^3)$
and hence has a strong dependence on the mass of the decaying quark. Similarly,
ratios of $B$-meson lifetimes can be recast,
without making any approximations, as:
\begin{eqnarray}
  \frac{\tau(B_q)}{\tau(B_{q^\prime})} & = & \frac{\Gamma_b + \delta \Gamma_{B_{q^\prime}}}{\Gamma_b + \delta \Gamma_{B_{q}}} =
  1 + \left( \delta \Gamma_{B_{q^\prime}} - \delta \Gamma_{B_q}\right)  \tau(B_q) \, .
  \label{eq:ratio1}
\end{eqnarray}
In our analysis we obtain predictions for the lifetime ratios by combining the HQE result for
$ (\delta \Gamma_{B_{q^\prime}} - \delta \Gamma_{B_q} )$ with the experimental value of $\tau(B_q)$, given the higher precision of the latter. Note that in this way, the numerically leading term  $\Gamma_b$ cancels out, so that $\tau(B_q)/\tau(B_{q^\prime})$
becomes sensitive only to subleading HQE corrections,
while the total decay rate can be used primarily to test our ability to predict the free-quark decay.
Alternatively, Eq.~\eqref{eq:ratio1} can be determined 
entirely within the HQE, in this case the dependence on $\Gamma_b$ is still present albeit very mildly, leading to slightly larger uncertainties.
\\
In the presence of physics beyond the SM (BSM), lifetime ratios will be modified as:
\begin{eqnarray}
  \frac{\tau(B_q)}{\tau(B_{q^\prime})}  & = &  
   1 + \left( \delta \Gamma_{B_{q^\prime}}^{\rm SM} - \delta \Gamma_{B_q}^{\rm SM} \right) \tau(B_q)
  \nonumber
  \\
  & + & \Br (B_{q^\prime} \to X )^{\rm BSM}   \frac{\tau (B_q)}{ \tau(B_{q^\prime})} - \Br (B_q \to Y)^{\rm BSM} .
\end{eqnarray}
For the case of the ratio $\tau (B_s) / \tau (B_d)$, a precision of the per mille level - both in experiments and theory - will allow
indirect new physics searches $B_q \to X$ at the same level of accuracy. 
However, also a precision of the order of one or two per cent in the theory prediction of 
$\tau (B^+) / \tau (B_d)$ can give interesting constraints on certain BSM models.
Examples of new $B_q \to X$ transitions that can modify the predictions for the $B$-meson lifetimes are:
\begin{itemize}
\item BSM contributions to non-leptonic tree-level $b$-quark decays, like $b \to c \bar{u} d$, 
$b \to c \bar{u} s$
      or $b \to c \bar{c} s$ transitions, see e.g. Refs.~\cite{Bobeth:2014rda,Brod:2014bfa, Jager:2017gal,Jager:2019bgk,Lenz:2019lvd,Bordone:2020gao,Iguro:2020ndk,Cai:2021mlt}.
\item BSM effects to $b \to s \tau \tau$  
          transitions, see e.g.\ Ref.~\cite{Bobeth:2011st} -
          such enhanced contributions could stem from models that explain the $b \to s \ell \ell$
          anomalies, see e.g.\ Refs.~\cite{Kamenik:2017ghi,Capdevila:2017iqn,Cornella:2020aoq}. 
          Moreover, direct bounds on this channel
          are very weak \cite{LHCb:2017myy}, and in the range of the experimental precision of the lifetime ratio $\tau (B_s)/\tau (B_d)$:
            \begin{eqnarray}
              \Br( B_s \to \tau^+ \tau^-) & < & 6.8 \cdot 10^{-3} \, \, \, \mbox{ (LHCb)} \, .
                     \end{eqnarray}
            The corresponding SM prediction \cite{Bobeth:2013uxa} lies far below
            the current experimental bound:
            \begin{eqnarray}
              \Br( B_s \to \tau^+ \tau^-)^{\rm SM}   & = & (7.73 \pm 0.49)  \cdot 10^{-7}  \, .
            \end{eqnarray}
\item BSM contributions that arise in the baryogenesis models presented in Refs.~\cite{Elor:2018twp,Alonso-Alvarez:2021qfd} 
contain new $b$-quark decay channels that affect the lifetimes.
\end{itemize}
We also point out that a precise determination of $\tau (B_s)/ \tau (B_d)$ and $\tau (B^+) / \tau (B_d)$ can be further used to constrain the possible size of duality
violating effects in the theoretical determination of lifetimes, see Ref.~\cite{Jubb:2016mvq}.
\\

\noindent
In light of the increasing experimental precision and of the phenomenological potential outlined above, we present an update of the SM prediction for $B$-meson lifetimes. Specifically, our study includes the following improvements with respect to previous works:
\begin{enumerate}
    \item 
    The Wilson coefficient of the Darwin operator. This represents a correction of order $1/m_b^3$ to the HQE and only recently the corresponding expressions for non-leptonic $b$-quark decays have been computed in Refs.~\cite{Lenz:2020oce, Mannel:2020fts, Moreno:2020rmk, Piscopo:2021ogu}. Thus, in the present work we include for the first time the complete dimension-six contribution at LO-QCD. Interestingly, the Darwin operator leads to sizeable effects to the lifetimes and in particular constitutes   one of the dominant contributions (or even the dominant one depending on the input for the matrix elements and SU(3)$_F$ breaking) to the lifetime ratio $\tsd$.
    
    \item 
    SU(3)$_F$ breaking corrections to the matrix elements of dimension-six four-quark operators as
    recently computed in Ref.~\cite{King:2021jsq}. 
    These matrix elements were first determined with heavy quark effective theory (HQET) sum rules in
    Ref.~\cite{Kirk:2017juj} for the case of the $B_d$ meson.   
    It is worthwhile to stress that so far no lattice determination 
    is available in the literature, the most recent estimates date back to  proceedings from 2001 \cite{Becirevic:2001fy}, while the corresponding publication has never appeared. 
     
    \item 
    Consistent determination of  dimension-seven four-quark operator contributions. In fact, as it has been pointed out in Ref.~\cite{charm}, previous studies were incorrectly including the effect of some dimension-seven operators, which was actually already accounted for when converting the HQET decay constant to the QCD one.
    
    \item 
    The so-called eye-contractions, which have been computed for the first time in Ref.~\cite{King:2021jsq} with HQET sum rules.~In this approach the eye-contractions constitute subleading corrections of order $\alpha_s$ to the matrix element of dimension-six four-quark operators and their numerical effect is found to be small. 
    
    \item 
    Detailed numerical analysis of the total decay rates of the $B_d$ , $B^+$ and $B_s$ mesons.
    
    \item 
    Update of all relevant SM parameters, in particular the CKM input.
    
    \end{enumerate}

\noindent
The paper is structured as follows: in Section~\ref{sec:Th-Fram} we present the main ingredients of the analysis. Specifically, in Section~\ref{sec:Eff-Ham} we outline the general structure of the HQE for the $b$-system, in Section~\ref{sec:Short-dis} we describe the status and the updates for the short-distance contributions, while in Section~\ref{sec:Non-pert-input}
 we analyse in detail the non-perturbative part of the HQE and the choice of the corresponding input.
 Our numerical results are discussed in
 Section~\ref{sec:Res} and we conclude with a summary and
 an outlook in Section~\ref{sec:Conc}. Finally, all inputs
 used in our analysis are collected in
 Appendix~\ref{app:1}, and we provide the complete
 expressions for the contribution of dimension-seven
 four-quark operators in HQET in Appendix~\ref{app:2}.

\section{Theoretical framework}
\label{sec:Th-Fram}
\subsection{Effective Hamiltonian and HQE}
\label{sec:Eff-Ham}
The most general effective Hamiltonian  describing the weak decays of a $b$-quark, see e.g. Ref.~\cite{Buchalla:1995vs}, takes the schematic form:
\begin{equation}
  {\cal H}_{\rm eff}  =   {\cal H}_{\rm eff}^{\rm NL} + {\cal H}_{\rm eff}^{\rm SL} + {\cal H}_{\rm eff}^{\rm rare} \,.
\label{eq:Heff-complete}
\end{equation}
In the above equation, ${\cal H}_{\rm eff}^{\rm NL}$ indicates the contribution due to non-leptonic $b$-quark transitions:
\begin{align}
  {\cal H}_{\rm eff}^{\rm NL} = 
  \frac{G_F}{\sqrt{2}} \sum_{q_3 = d, s}
  \left[\,
   \sum_{\substack{q_{1,2} = u, c} } \!\! \lambda_{q_1 q_2 q_3} 
  \Bigl(C_1 (\mu_1) \, Q_1^{q_1 q_2 q_3}  + C_2 (\mu_1) \, Q_2^{q_1 q_2 q_3}  \Bigr)
    -  \lambda_{q_3} 
  \!\! \!\! \sum \limits_{j=3, \ldots, 6, 8} \! \!\! C_j (\mu_1) \, Q_j^{q_3} 
   \right] + {\rm h.c.}\, ,
   \label{eq:Heff-NL}
\end{align}
where $\lambda_{q_1 q_2 q_3} = V_{q_1 b}^* V_{q_2 q_3} $
and $\lambda_{q_3} = V_{tb}^* V_{tq_3} $ stand for the corresponding CKM factors, 
$C_i (\mu_1)$ denote the Wilson coefficients of the $\Delta B = 1$ effective operators evaluated at the renormalisation scale $\mu_1 \sim m_b$, while
$Q_{1,2}^{q_1 q_2 q_3}$, $Q_j^{q_3}$ ($j = 3, \ldots, 6$) and $Q_8^q$ indicate respectively the current-current~\footnote{We emphasise that in our notation $Q_1^{q_1q_2 q_3}$ is the colour-singlet operator, contrary to e.g.\ Ref.~\cite{Buchalla:1995vs}.}, the penguin and the chromo-magnetic operators. They have the following expressions:
\begin{equation}
Q_1^{q_1 q_2 q_3} 
 =   
\left(\bar b^i \, \Gamma_\mu \, q_1^i \right)
\left(\bar q_2 \, \Gamma^\mu  \, q_3^j \right)\,,
\qquad 
Q_2^{q_1 q_2 q_3} 
 =  
\left(\bar b^i  \, \Gamma_\mu  \, q_1^j \right)
\left(\bar{q}_2^j \, \Gamma^\mu  \, q_3^i \right)\,,
\label{eq:Q12}
\end{equation}
\begin{align}
Q_3^{q_3} 
& 
= (\bar b^i \, \Gamma_\mu \, q_3^i) \sum_{q} ( \bar q^j \, \Gamma^\mu \, q^j)
\,, \qquad 
Q_4^{q_3} = (\bar b^i \, \Gamma_\mu \, q_3^j) \sum_{q} (\bar q^j \, \Gamma^\mu \, q^i)\,, 
\label{eq:Q34}
\\
Q_5^{q_3} 
& 
=  (\bar b^i \, \Gamma_\mu \, q_3^i) \sum_{q} (\bar q^j \, \Gamma_+^\mu \, q^j)\,, 
\qquad
Q_6^{q_3} = (\bar b^i \, \Gamma_\mu \, q_3^j) \sum_{q} 
(\bar q^j \, \Gamma_+^\mu \, q^i)\,,
\label{eq:Q56}
\end{align}
\begin{equation}
Q_8^{q_3} = \frac{g_s}{8 \pi^2} m_b
\left(\bar b^i \, \sigma^{\mu\nu} (1 - \gamma_5) t^a_{ij} \, q_3^j \right) G^a_{\mu \nu}\,,
\label{eq:Q8}
\end{equation}
with $\Gamma_\mu = \gamma_\mu(1-\gamma_5)$,  $\Gamma_+^\mu = \gamma^\mu(1+\gamma_5)$ and $\sigma_{\mu \nu} =(i/2) [\gamma_\mu, \gamma_\nu]$. Moreover, in Eqs.~(\ref{eq:Q12}) - (\ref{eq:Q8}), $g_s$~denotes the strong coupling, $G_{\mu\nu} = G^a_{\mu\nu}t^a $ is the gluon field strength tensor, while $i,j = 1, 2, 3$ and $a = 1, \ldots, 8$ label the $SU(3)_c$ indices for fields respectively in the fundamental and in the adjoint representation. A comparison of the 
values of the corresponding Wilson coefficients for different choices of the scale $\mu_1$ and both at LO- and NLO-QCD \cite{Buchalla:1995vs} is shown in the Appendix in Table~\ref{tab:WCs}.
The second term in Eq.~\eqref{eq:Heff-complete} describes the contribution to the effective Hamiltonian due to semileptonic $b$-quark decays, i.e.
\begin{equation}
{\cal H}_{\rm eff}^{\rm SL} 
= 
\frac{G_F}{\sqrt 2} \sum_{q = u, c \,} \sum_{\, \ell = e, \mu, \tau}
V_{qb}^* \, Q^{q \ell} + {\rm h.c.}\,,
\label{eq:Heff-SL}
\end{equation}
with 
\begin{equation}
    Q^{q \ell} =\left(\bar{b}\, \Gamma^\mu \, q \right)
\left(\bar \nu_\ell \, \Gamma_\mu \, \ell \right)\,.
\end{equation}
Finally, ${\cal H}_{\rm eff}^{\rm rare}$ in Eq.~\eqref{eq:Heff-NL} encodes the contribution due to suppressed $b$-quark transitions which are only relevant for the study of rare decays like $B \to K^{(*)} \gamma$ or $B \to K \ell^+ \ell^-$. These modes have very small branching fractions which fall out of the current theoretical sensitivity for the lifetimes~\footnote{E.g.\ the
inclusive radiative decay $B \to X_s \gamma$  
has branching fractions of the order of $10^{-4}$, which is considerably below the current accuracy of our analysis.}. Hence, in the following, the effect of ${\cal H}_{\rm eff}^{\rm rare}$ is neglected and for brevity we do not show its explicit expression below.\\
\noindent
The total decay width of a $B_q$ meson, with mass~$m_{B_q}$ and four-momentum $p_{B}$ reads  
\begin{equation}
\Gamma ({B_q})  = 
\frac{1}{2 m_{{B_q}}} \sum_{X}  \int \limits_{\rm PS} (2 \pi)^4  \delta^{(4)}(p_{B}- p_X) \, \,
|\langle X(p_X)| {\cal H}_{\rm eff} | B_q(p_{B}) \rangle |^2,
\label{eq:Gamma-D}
\end{equation}
where PS~denotes the phase space integration, and the summation over all possible final states $X$ into which the $B$ meson can decay is performed. Using the optical theorem, $\Gamma(B_q)$  can be related to the discontinuity of the forward scattering matrix element of the time ordered product of the double insertion of the effective Hamiltonian, i.e.
\begin{equation}
\Gamma (B_q) =    \frac{1}{2 m_{B_q}} {\rm Im}
\langle B_q | {\cal T}| B_q \rangle \, ,
\label{eq:Gamma_opt_th}
\end{equation}
with the transition operator given by
\begin{equation}
{\cal T}  =  
i \int d^4x 
\,  T \left\{ {\cal H} _{\rm eff} (x) \, ,
 {\cal H} _{\rm eff} (0)  \right\} \, .
 \label{eq:optical_theorem}
\end{equation}
The non-local operator in Eq.~\eqref{eq:optical_theorem} can be evaluated by exploiting the fact that the $b$-quark is heavy i.e.\ $m_b \gg \Lambda_{\rm QCD}$, where the latter defines a typical hadronic scale. In fact, in the framework of the HQE, the $b$-quark momentum is  decomposed as
\begin{equation}
p_b^\mu = m_b  v^\mu + k^\mu\,,
\label{eq:c-quark-momentum}
\end{equation}
with $v = p_B/m_{B_q}$ denoting the four-velocity of the $B$-meson, and $k$  representing a residual momentum which accounts for non-perturbative
interactions of the $b$-quark with the light degrees of freedom, i.e.\ soft gluons and quarks, inside the hadronic state. It thus follows that $k \sim \Lambda_{\rm QCD}$. 
Moreover, the heavy $b$-quark field is parametrised as  
\begin{equation}
b (x) = e^{ - i m_b v \cdot x} b_v (x)\,,  
\label{eq:phase-redef}    
\end{equation}
by factoring out the large component of its momentum and by introducing a rescaled field $b_v(x)$ containing only low oscillation  frequencies of the order of $k$. In fact, this field  satisfies
\begin{equation}
    i D_\mu b(x) = e^{- i m_b v \cdot x} (m_b v_\mu + i D_\mu) b_v(x)\,,
    \label{eq:iDmu-b}
\end{equation}
showing that the action of the covariant derivative $D_\mu = \partial_\mu - i g_s A_\mu^a \, t^a $ also leads to a large contribution proportional to the heavy quark mass and to a residual term of the order of $\Lambda_{\rm QCD}$. Moreover, we recall that the rescaled field $b_v(x)$ is related to HQET field $h_v(x)$, see e.g.\ Ref.~\cite{Neubert:1993mb}, by 
\begin{equation}
b_v (x) = h_v (x) + \frac{i \slashed D_\perp}{2 m_b} h_v (x)  
+ {\cal O} \left(\frac{1}{m_b^2} \right)\,,
\label{eq:bv-hv-relation}
\end{equation}
with $D_\perp^\mu = D^\mu - (v \cdot D) \, v^\mu$.
\begin{figure}[t]
    \centering
    \includegraphics[scale=0.36]{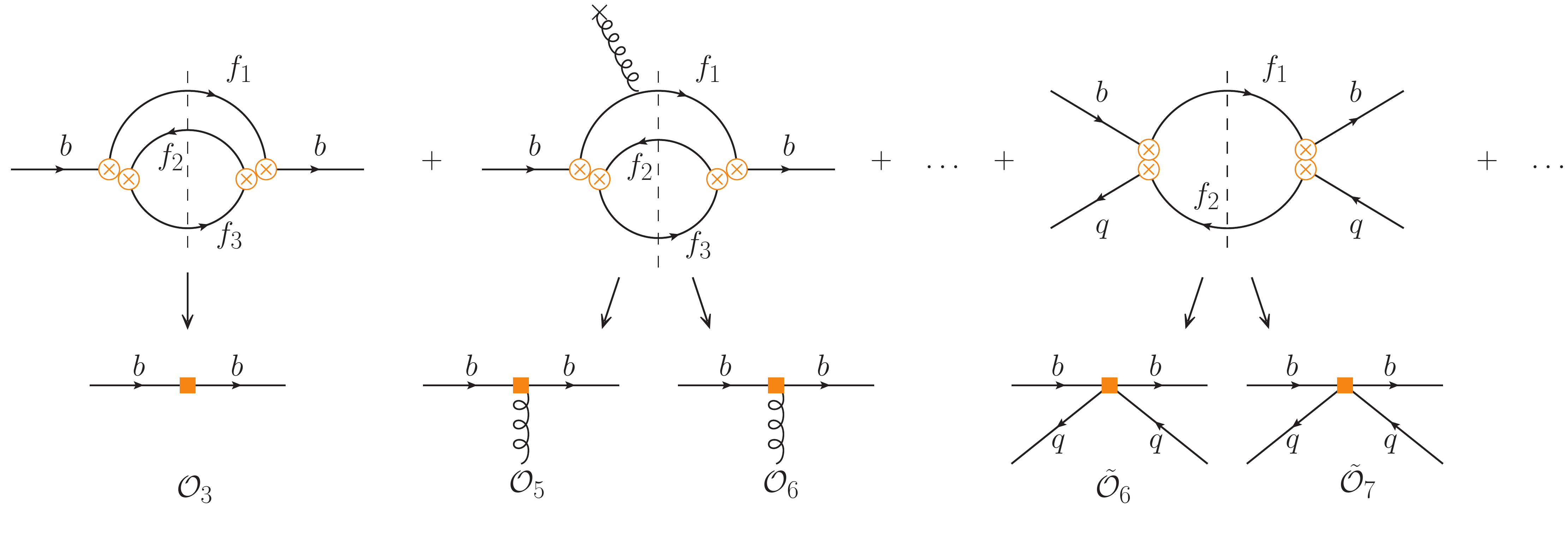}
    \caption{Schematic representation of the HQE for the total decay width of a $B$-meson.  
    The crossed circles and the full squares denote respectively the insertion of the $ \Delta  B = 1$
operators $Q_i$ of the effective Hamiltonian, and of the $\Delta B = 0$ operators ${\cal O}_i$ and
$\tilde{{\cal O}_i}$ in the HQE. Note that while the contribution of two-quark operators derives from two-loop diagrams, four-quark operators are generated already at one-loop at LO-QCD. The labels $f_{1,2,3}$ stand for all the possible fermions the $b$-quark can decay into.}
    \label{fig:HQE-scheme}
\end{figure}
Taking into account Eqs.~\eqref{eq:c-quark-momentum} - \eqref{eq:iDmu-b}, the total decay width in Eq.~\eqref{eq:Gamma_opt_th} can be systematically expanded in inverse powers of the heavy quark mass, 
leading to the HQE series. This schematically reads
\begin{equation}
\Gamma(B_q) = 
\Gamma_3  +
\Gamma_5 \frac{\langle {\cal O}_5 \rangle}{m_b^2} + 
\Gamma_6 \frac{\langle {\cal O}_6 \rangle}{m_b^3} + ...  
 + 16 \pi^2 
\left( 
  \tilde{\Gamma}_6 \frac{\langle \tilde{\mathcal{O}}_6 \rangle}{m_b^3} 
+ \tilde{\Gamma}_7 \frac{\langle \tilde{\mathcal{O}}_7 \rangle}{m_b^4} + ... 
\right),
\label{eq:HQE}
\end{equation}
where $\Gamma_i$ are short-distance functions which can be computed perturbatively in QCD, i.e.
\begin{equation}
\Gamma_i = \Gamma_i^{(0)} + \frac{\alpha_s}{4 \pi} \Gamma_i^{(1)} 
+ \left(\frac{\alpha_s}{4 \pi}\right)^2 \Gamma_i^{(2)}+ \ldots \, ,  
\label{eq:Gamma-i-pert-series}
\end{equation}
and $\langle {\cal O}_i \rangle \equiv
\langle B_q | {\cal O}_i |B_q  \rangle/(2 m_{B_q})$ denote the matrix elements of the corresponding $\Delta B = 0$ operators 
${\cal O}_i$ in the effective theory. Note that at the same order in $1/m_b$, both two- and four-quark operator contributions appear. The latter originate from loop-enhanced diagrams, as reflected by the explicit factor of $16 \pi^2$ in Eq.~\eqref{eq:HQE}, and to avoid confusion in the notation, we use a tilde to label them. The diagrammatic representation of 
Eq.~\eqref{eq:HQE} is shown in Fig.~\ref{fig:HQE-scheme}.

\subsection{Short-distance contributions}
\label{sec:Short-dis}
\begin{figure}[t]\centering
\includegraphics[scale=0.45]{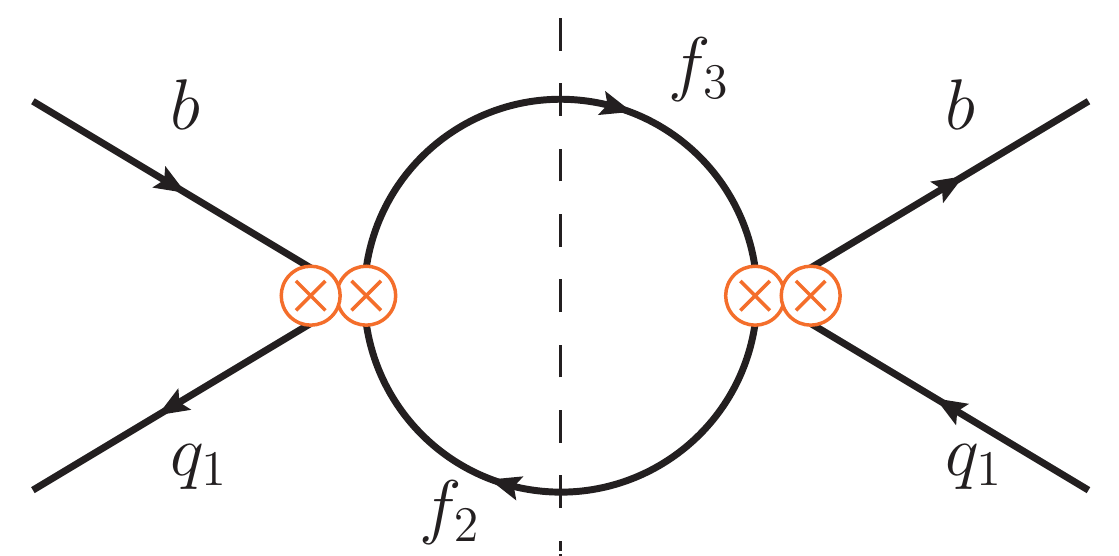}
\qquad
\includegraphics[scale=0.45]{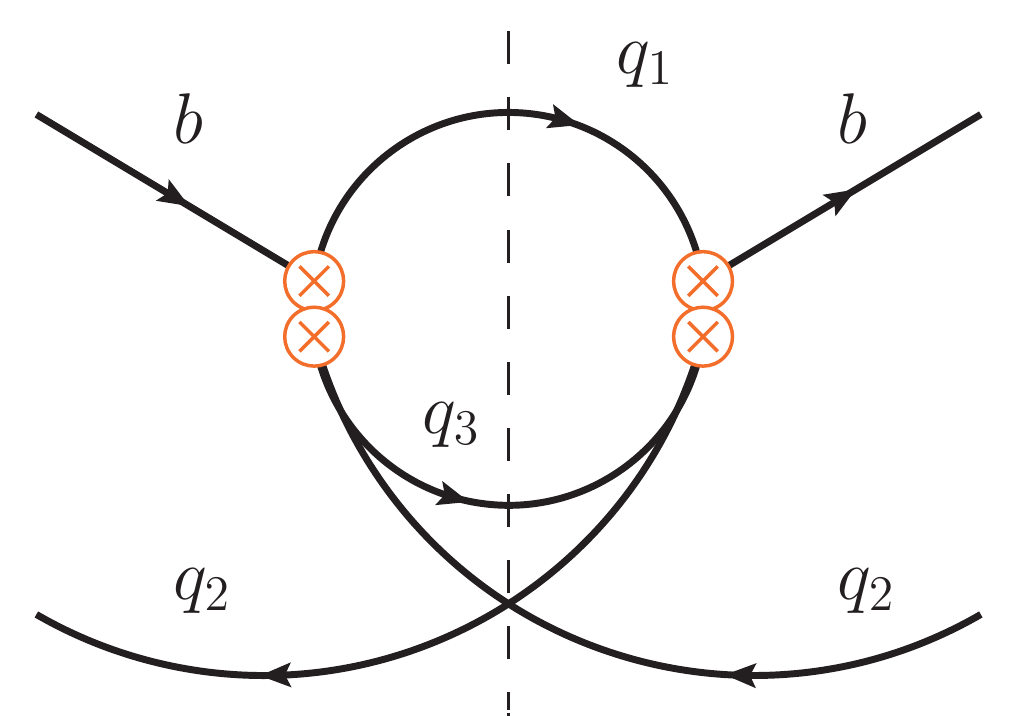}
\qquad
\includegraphics[scale=0.45]{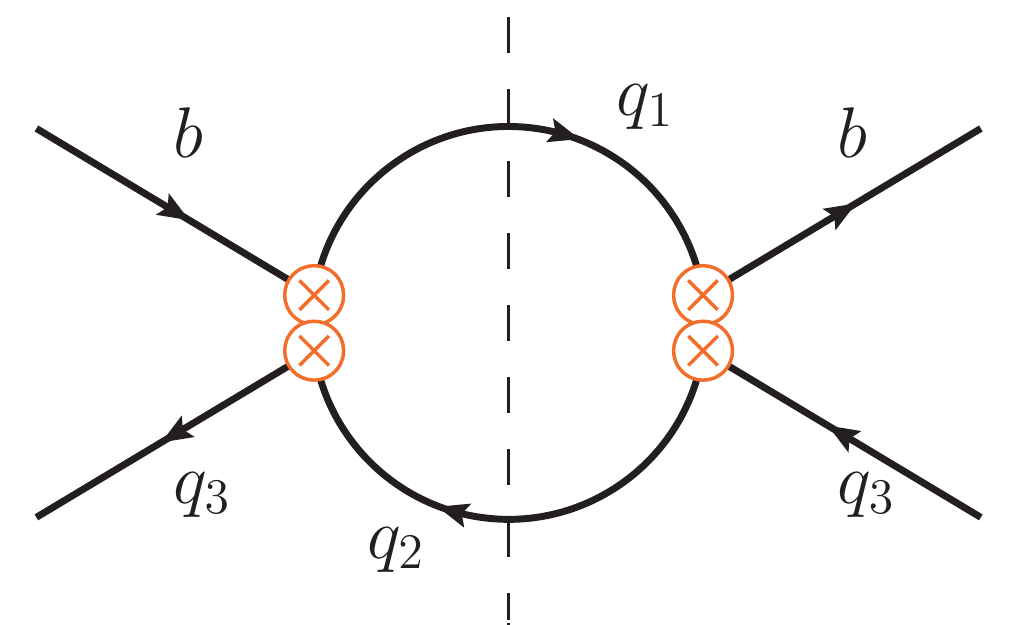}
\caption{Diagrams corresponding, from left to right, to the WA, PI and WE~topology.}
\label{fig:PI-WE-WA}
\end{figure}

In this section, we give a brief summary of the short-distance
contributions, cf.\ Eq.~\eqref{eq:Gamma-i-pert-series}, included in our analysis; however, for more details we refer to the recent study \cite{charm}, where a comprehensive description of the structure of the HQE for the charm sector has been discussed.
Note, that the coefficients $\Gamma_i, \tilde \Gamma_i$ are analytic functions of the masses of the internal fermions, and in our case, since we only keep non-vanishing the masses of the charm quark and of the tau-lepton
\footnote{ Since $m_s^2/m_b^2 \approx m_\mu^2/m_b^2 \sim 0.05 \%$, the effect of non-vanishing $s$-quark and $\mu$-lepton masses to the short-distance coefficients is far below the current theoretical accuracy and hence can be safely neglected. However, we must include strange quark mass corrections in the non-perturbative input, where these effects are much more pronounced, in order to account for 
SU(3)$_F$ breaking.
}, they depend on the two dimensionless mass parameters:
\begin{equation}
    \rho = \frac{m_c^2}{m_b^2} \,, \qquad \eta = \frac{m_\tau^2}{m_b^2} \,.
\end{equation}
The leading contribution to the $B$-meson total decay width corresponds to the free $b$-quark decay, obtained by computing the discontinuity, at LO-QCD, of the two-loop diagrams schematically pictured on the top left of Fig.~\ref{fig:HQE-scheme}. 
A compact expression for this coefficient takes the form:
\begin{equation}
\Gamma_3 = \Gamma_0 \, c_3 = \Gamma_0 \left( c_3^{(0)} + \frac{\alpha_s}{4 \pi} c_3^{(1)} + \ldots \right)\,,
\end{equation}
where
\begin{equation}
\Gamma_0 = \frac{G_F^2 \, m_b^5}{192 \pi^3} 
|V_{cb}|^2\,,
\label{eq:Gamma0}
\end{equation}
and 
\begin{equation}
    c_3 = 
     {\cal C}_{3, \rm SL}  +
    3 \, C_1^2    \, {\cal C}_{3,11} 
 +  2 \, C_1 C_2  \, {\cal C}_{3,12} 
 +  3 \, C_2^2    \, {\cal C}_{3,22}
 + C_i \, C_j \, {\cal C}^{P}_{3, ij} 
 \,.
 \label{eq:c3-decomposition}
\end{equation}
In Eq.~\eqref{eq:c3-decomposition}, a summation over all possible non-leptonic and semileptonic modes of the $b$-quark is implicitly assumed and we have denoted by ${\cal C}_{3, ij}^P$ with $i = 1, 2$ and $j = 3, \ldots, 6, 8$ the contribution due to the mixed insertion of the current-current and of the penguin or the chromo-magnetic operators. Remarkably, for semileptonic modes even $\alpha_s^3$ corrections have been computed by now \cite{Fael:2020tow, Czakon:2021ybq}, however, as the accuracy for non-leptonic modes reaches only NLO-QCD, we perform our analysis consistently at this order and do not include the new results for ${\cal C}_{3, {\rm SL}}$. Moreover, following a common counting adopted in the literature \cite{Lenz:1997aa, Lenz:1998qp}, due to the small size of the corresponding Wilson coefficients, the contribution of the penguin and chromo-magnetic operators is in fact treated as a next-to-leading order effect, i.e. ${\cal C}^P_{3, ij} = 0$, for $j = 3, \ldots, 6, 8$
at LO-QCD.
The analytic expressions for 
${\cal C}_{3, 11}$, ${\cal C}_{3, 22}$, and ${\cal C}_{3, {\rm SL}}$ can be easily extracted from Ref.~\cite{Hokim:1983yt}, 
where the computation has been performed for three different final state masses, 
while those for ${\cal C}_{3, 12}$ are derived from the results presented in Ref.~\cite{Krinner:2013cja} in the case of the $b \to c \bar c s$ transition, and in Ref.~\cite{Bagan:1994zd} for the remaining modes. Finally ${\cal C}_{3, ij}^P$ are taken from Ref.~\cite{Krinner:2013cja}.\\
At order $1/m_b^2$, the short-distance coefficients are obtained by computing the discontinuity of two-loop diagrams, as the one in the center of Fig.~\ref{fig:HQE-scheme}, and by taking into account the expansion of the dimension-three matrix element. The corresponding contribution can be schematically written as 
\begin{equation}
    \Gamma_5 \frac{\langle {\cal O}_5 \rangle}{m_b^2} 
    = \Gamma_0 
    \left[
c_{\pi} \frac{\langle {\cal O}_{\pi}\rangle}{m_b^2} 
    + c_G \, \frac{\langle {\cal O}_{G}\rangle}{m_b^2} 
    \right]\,,
    \label{eq:Gamma_5}
\end{equation}
where the matrix elements of the kinetic and chromo-magnetic operators, see Eqs.~\eqref{eq:Okin}, \eqref{eq:Omag} \footnote{Note that with a little abuse of notation, we denote by chromo-magnetic operators both $Q^{q_3}_8$ and ${\cal O}_G$. However, as they arise respectively in the $\Delta B = 1$ and $\Delta B = 0$ effective theory, it should be clear from the context to which one we actually refer.}, are  discussed in Section \ref{sec:Non-pert-input}.
In our analysis, again for consistency, we include the coefficients $c_{\pi}$ and 
$c_{G}$ only at LO-QCD, since $\alpha_s$ corrections have so far been determined only for the semileptonic channels~\cite{Mannel:2015jka}. At this order, 
the contribution of the kinetic operator is equal to the one of dimension-three up to a numerical factor, i.e.\
$c_{\pi}= - c_3^{(0)}/2$,
while the coefficient $c_G$ can be decomposed as
\begin{equation}
    c_G = 
     {\cal C}_{G,\rm SL}+
    3 \, C_1^2    \, {\cal C}_{G,11} 
 +  2 \, C_1 C_2  \, {\cal C}_{G,12} 
 +  3 \, C_2^2    \, {\cal C}_{G,22} \, ,
\label{eq:CG}
\end{equation}
here again a summation over all the $b$-quark modes is assumed.
The expressions for the non-leptonic channels  ${\cal C}_{G,ij}$ can be found e.g.\ in the Appendix of Ref.~\cite{Lenz:2020oce}, however originally computed in Refs.~\cite{Blok:1992he, Blok:1992hw, Bigi:1992ne}, while
the semileptonic coefficient ${\cal C}_{G, SL}$ is obtained using the general result for two different final state masses presented e.g.\ in the Appendix of Ref.~\cite{Mannel:2017jfk},
and first determined in Refs.~\cite{Balk:1993sz, Falk:1994gw}.
\\ 
At order $1/m_b^3$, both
two- and four-quark operators contribute \footnote{We stress that, by using the equations of motion for the gluon field strength tensor, the dimension-six operator basis can be also written in terms of four-quark operators only, see Section \ref{sec:Non-pert-input}.}, see respectively 
the second and third diagram on the top line of Fig.~\ref{fig:HQE-scheme}. For the former, we can compactly write
\begin{eqnarray}
\Gamma_6 \frac{\langle {\cal O}_6 \rangle}{m_b^3}
=
\Gamma_0 \, c_{\rho_D} \frac{\langle {\cal O}_{D}\rangle}{m_b^3} \, ,
\label{eq:Gamma_6}
\end{eqnarray}
where the matrix element of the Darwin operator is defined in Eq.~\eqref{eq:OD}, while the corresponding short-distance coefficient can be decomposed as:
\begin{equation}
    c_{\rho_D} = 
    {\cal C}_{\rho_D,\rm SL} 
    +
    3 \, C_1^2 \, {\cal C}_{\rho_D,11} 
 +  2 \, C_1 C_2 \, {\cal C}_{\rho_D,12} 
 +  3 \, C_2^2 \, {\cal C}_{\rho_D,22}  \,,
\label{eq:crhoD}
\end{equation}
summing again over all $b$-quark decay modes. Also in this case, the accuracy in our analysis reaches only LO-QCD, as this is the order at which the non-leptonic contributions are known; for the semileptonic decays the coefficient of the Darwin operator has been first computed in Ref.~\cite{Gremm:1996df},
while NLO-QCD corrections have been recently determined in Ref.~\cite{Mannel:2019qel, Mannel:2021zzr, Moreno:2022goo}.
The complete expressions of ${\cal C}_{\rho_D, ij}$ for all non-leptonic channels have been obtained recently in Refs.~\cite{Lenz:2020oce, Mannel:2020fts, Moreno:2020rmk}, while the coefficient ${\cal C}_{\rho_D, {SL}}$ can be read off the general results for the case of two different final state masses presented e.g. in Refs.~\cite{Rahimi:2022vlv, Moreno:2022goo}.
 In this respect, it is worth emphasising that contrary to the naive expectation, 
the coefficient of the Darwin operator is found to be sizeable; more precisely, it results to be one order of magnitude larger that the corresponding ones at dimension-five.
However, as it has been shown in detail e.g.\ in Ref.~\cite{Lenz:2020oce}, this  actually follows from an accidental suppression of the dimension-five coefficients, and not from an abnormal enhancement of the Darwin term.
Therefore, the contribution of the Darwin operator,  
neglected in previous phenomenological studies, turns out to be an important ingredient for the theoretical prediction of the $B$-meson lifetimes, see  Section~\ref{sec:Res}.\\
The short-distance coefficients due to four-quark operators are obtained by computing, at LO-QCD, the discontinuity of the one-loop diagrams showed in Fig.~\ref{fig:PI-WE-WA}, corresponding respectively to the weak annihilation (WA), Pauli interference (PI), and weak-exchange (WE) topologies, see e.g. Refs.~\cite{Uraltsev:1996ta,Neubert:1996we} for results including the charm quark mass dependence.
Their contribution, at dimension-six,
can be schematically written as
\begin{align}
16 \pi^2 \, \tilde{\Gamma}_6 \frac{\langle \tilde{\cal O}_6\rangle}{m_b^3}
 = 
\Gamma_0 
\Biggr[
A_{i, q_1 q_2}^{\rm WE} \frac{\langle {\tilde O}_i^{q_3} \rangle}{m_b^3}
+ A_{i, q_1 q_3}^{\rm PI} \frac{\langle {\tilde O}_i^{q_2}\rangle}{m_b^3}
+ \, 
A_{i, q_2 q_3}^{\rm WA} \frac{\langle {\tilde  O}_i^{q_1} \rangle}{m_b^3}
+
A_{i, q_1 \ell}^{\rm WA} 
\frac{\langle {\tilde  O}_i^{q_1} \rangle}{m_b^3}
\Biggl]\,,
\label{eq:dim-6-4q-NLO-scheme}
\end{align}
where $i = 1, \ldots, 4$ and a sum over all possible final states, following the notation in Eq.~\eqref{eq:Heff-NL}, is implied. Again, we refer to Section~\ref{sec:Non-pert-input} for a discussion of the matrix elements of the corresponding four-quark operators. Moreover, 
 Eq.~\eqref{eq:dim-6-4q-NLO-scheme} shows that, contrary
 to the corrections described so far, now differences in
 the contributions to specific $B_q$-mesons arise not
 only because of different states in the matrix elements,
 but also due to different short-distance coefficients.
 In light of this and of the formal loop enhancement with
 respect to two-quark operators, the effect of four-quark
 operators was expected to give the dominant
 correction to the total widths and in particular to the
 lifetime ratios, see e.g. Refs.~\cite{Uraltsev:1996ta,Neubert:1996we}. 
The complete expressions for $A_{i, q_1 q_2}^{\rm WE}$ and $A_{i, q_1 q_3}^{\rm PI}$ up to NLO-QCD corrections, and including also the effect of mixed tree-penguin contributions,
have been
computed in Ref.~\cite{Franco:2002fc} for four-quark operators defined in HQET\footnote{For the case of QCD operators see
Refs.~\cite{Franco:2002fc,Beneke:2002rj}.}.
The results for $A_{i, q_2 q_3}^{\rm WA}$ can be
obtained, by means of a Fierz transformation,  
from the corresponding ones for 
$A_{i, q_1 q_2}^{\rm WE}$ replacing  $C_1 \leftrightarrow
C_2$. For semileptonic modes, the coefficients
$A_{i, q_1 \ell}^{\rm WA}$ have been determined in Ref.~\cite{Lenz:2013aua}.\\
Finally, at order $1/m_b^4$, only the LO-QCD short-distance coefficients 
of the four-quark operators are known in the literature. 
They were determined in Refs.~\cite{Gabbiani:2003pq, Gabbiani:2004tp} 
for operators defined in QCD\footnote{We note that while comparing 
our results with the ones presented 
in Refs.~\cite{Gabbiani:2003pq, Gabbiani:2004tp}, 
we have actually found some inconsistencies in their expressions 
and communicated this to the authors.} 
and also in Ref.~\cite{Lenz:2013aua} for the HQET ones.
The corresponding contribution to the total decay width schematically reads
\begin{align}
16 \pi^2 \, \tilde{\Gamma}_7 \frac{\langle \tilde{\cal O}_7\rangle}{m_b^4}
& =  
\Gamma_0 
\Biggr[
B_{i, q_1 q_2}^{\rm WE} \frac{\langle {\tilde P}_i^{q_3} \rangle}{m_b^4}
+ B_{i, q_1 q_3}^{\rm PI} \frac{\langle {\tilde P}_i^{q_2}\rangle}{m_b^4}
+ \, 
B_{i, q_2 q_3}^{\rm WA} \frac{\langle {\tilde  P}_i^{q_1} \rangle}{m_b^4}
+
B_{i, q_1 \ell}^{\rm WA} 
\frac{\langle {\tilde  P}_i^{q_1} \rangle}{m_b^4}
\Biggl]\,,
\label{eq:dim-7-4q-LO-scheme}
\end{align}
where $i = 1, \ldots, 18$. 
We refer to Appedix~\ref{app:2} for the analytic expressions of the dimension-seven four-quark operator contribution to the WE, PI, and WA topologies in HQET. 
The dimension-seven four-quark operator basis, together with the relative parametrisation, is briefly discussed in Section~\ref{sec:Non-pert-input}, while more details can be found in Ref.~\cite{charm}.

\subsection{Non-perturbative input}
\label{sec:Non-pert-input}
In this section we discuss the status and the choice of the non-perturbative input needed in our analysis. These parametrise the matrix elements of the $\Delta B = 0$ operators in the HQE, and hence encode the contribution to the total decay width due to hadronic effects.
\\
For clarity of the presentation, it is convenient to start with the four-quark operators.
At order $1/m_b^3$ in the HQE, we consider the following operator basis, already defined in terms of the HQET field $h_v(x) = b_v(x) + {\cal O} (1/m_b)$:
\begin{align}
  {\tilde O}_1^q  
  & =  
  (\bar{h}_v\,\gamma_\mu (1-\gamma_5) q)\,(\bar{q}\,\gamma^\mu (1-\gamma_5) h_v) ,
  \label{eq:O1-HQET} \\[1mm]
  {\tilde O}_2^q  
  & =  
  (\bar{h}_v (1 - \gamma_5) q)\,(\bar{q} (1 + \gamma_5) h_v) ,
  \label{eq:O2-HQET} \\[1mm]
  {\tilde O}_3^q  
  & =  
  (\bar{h}_v \, \gamma_\mu (1-\gamma_5) \, t^a q) 
  \, (\bar{q} \, \gamma^\mu (1-\gamma_5) \, t^a  h_v) ,
  \label{eq:T1-HQET} \\[1mm]
  {\tilde O}_4^q 
  & =  
  (\bar{h}_v (1-\gamma_5) t^a q)\,(\bar{q}(1 + \gamma_5) t^a h_v).
  \label{eq:T2-HQET}
\end{align}
The matrix elements of the operators in Eqs.~\eqref{eq:O1-HQET} - \eqref{eq:T2-HQET} are parameterised as
\cite{King:2021jsq}
\begin{eqnarray}
\langle {B}_q | {\tilde O}_i^q \, | {B}_q \rangle 
& = & 
F_q^2(\mu_0) \, m_{B_q} \, \tilde B_i^q(\mu_0) \,,
\label{eq:ME-dim-6-HQET-q-q}
\\[2mm]
\langle {B}_q | {\tilde O}_i^{q^\prime} | {B}_q \rangle 
& = & 
F_q^2 (\mu_0) \, m_{B_q} \, \tilde \delta^{q^\prime q}_i (\mu_0)\,,
\qquad q \not = q^\prime \,,
\label{eq:ME-dim-6-HQET-q-q-prime}
\end{eqnarray}
where ${\tilde B}_i^q(\mu_0)$ \footnote{
Note that in the literature ${\tilde O}_{3,4}^q  $ are 
sometimes denoted by ${\tilde T}_{1,2}^q$, and 
correspondingly ${\tilde B}_{3,4}^q$ by 
$\epsilon_{1,2}^q$.} and $\tilde \delta^{ q^\prime q}_i(\mu_0)$, 
$i =1, ... 4$, denote, respectively, the Bag parameters and the so-called eye-contractions, evaluated at the 
renormalisation $\mu_0$. We emphasise that the 
eye-contractions correspond to subleading  effects of 
order $\alpha_s$ to the matrix elements, originating from diagrams in which the spectator quark in the $B$ meson 
and the light quark in the effective operator do not 
coincide \footnote{The eye-contractions with $q = q^\prime$
are in fact included in the Bag parameters $\tilde B_i^q$.}.
 Moreover, $F_q(\mu_0)$ in Eqs.~\eqref{eq:ME-dim-6-HQET-q-q} - \eqref{eq:ME-dim-6-HQET-q-q-prime} labels the HQET decay constant; this is related to the QCD decay constant $f_{B_q}$, at one-loop accuracy and up to power corrections, 
by~\cite{Neubert:1991sp, Neubert:1992fk} 
\begin{equation}
   f_{B_q} = \frac{F_q (\mu_0)}{\sqrt{m_{B_q}}} \left[1 +  
   \frac{\alpha_s(\mu_0)}{2\pi} 
   \left(\ln \left(\frac{m_b^2}{\mu_0^2} \right)
   - \frac 4 3 \right) + 
   {\cal O}\left(\frac{1}{m_b}\right) \right] \,.
   \label{eq:decay-const-conv}
\end{equation}
In vacuum insertion approximation (VIA), the Bag parameters of the colour-singlet operators are equal to one, while those of the colour-octet
operators and all the eye-contractions vanish:
\begin{equation}
    {\tilde B}_{1,2}^q \overset{\rm VIA}{=} 1\,, \qquad \tilde B^q_{3,4} \overset{\rm VIA}{=} 0\,, \qquad \tilde \delta^{q^\prime q}_i \overset{\rm VIA}{=} 0\,.
\end{equation}
The deviation of the non-perturbative input ${\tilde B}_i^q(\mu_0)$ and $ \tilde \delta_i^{q q^\prime}(\mu_0)$ from their VIA values
has been computed within the framework of HQET sum rules. Specifically,
the Bag parameters~${\tilde B}_i^q$ were firstly determined in Ref.~\cite{Kirk:2017juj} for the case of the $B^{+,0}$ 
mesons, while
corrections due to the strange quark mass as well as the
contribution of the eye-contractions
have been obtained in Ref.~\cite{King:2021jsq}.
Their numerical values are listed
in Table~\ref{tab:Bag-parameters}.
Note that throughout this work we assume isospin symmetry, i.e. 
\begin{equation}
{\tilde B}_i^u = {\tilde B}_i^d\,,
\qquad
\tilde \delta^{u s}_i = \tilde \delta^{d s}_i\,,
\qquad 
\tilde \delta^{s u}_i = \tilde \delta^{s d}_i\,,
\qquad 
\tilde \delta^{u d}_i = \tilde \delta^{d u}_i\,.
\end{equation}
At order $1/m_b^4$, the HQET operator basis is much 
larger, specifically it includes $18$ operators of which 
10 local and 8 non-local, obtained from the expansion of 
the heavy quark field and of the HQET Lagrangian, see 
e.g.\ Ref.~\cite{Neubert:1993mb}. Their complete 
expressions, together with the parametrisation of their 
matrix elements can be found in Ref.~\cite{charm}. Here 
we limit ourselves to stress that the effect of all 
non-local operators and of some local ones is actually 
absorbed in the conversion of the HQET decay constant, 
cf.\ Eq.~\eqref{eq:decay-const-conv}, when also $1/m_b$ 
corrections are taken into account~\footnote{
We stress that in principle the running of the dimension-seven HQET operators should also be taken into account and would lead to a residual effect not included in the QCD decay constant. A detailed study, performed in Ref.~\cite{Kilian:1992cj} for the case of $B - \bar B$ mixing, has found this effect to be small i.e.\  $\sim 5\%$ for running of $\mu_0$ from the scale $m_b$ to $1 \,{\rm GeV}$.
Since we consider $\mu_0 \sim m_b$ and do not run to low scales $\sim 1 \,{\rm GeV}$, we expect 
even a much smaller effect to the lifetimes 
and thus we neglect it in our analysis.}.
This has been overlooked in previous analyses  \cite{Lenz:2013aua} and only recently clarified in Ref.~\cite{charm}.
In Appendix~\ref{app:2}, we present the analytic expression for the dimension-seven contribution to the PI, WE and WA topologies in VIA. Note that so far, there is no computation of the dimension-seven Bag parameters available in the literature.
\\
We now turn to discuss the input needed to parametrise the matrix elements of two-quark operators. Up to order $1/m_b^3$ in the HQE, the  basis includes the kinetic and chromo-magnetic operators at dimension-five and the Darwin operator at dimension-six
\footnote{We stress that at dimension-six the basis formally includes also the spin-orbit operator ${\cal O}_{\rm LS}$. However, by adopting definitions in terms of full covariant derivatives (and not the transversal ones), the contribution of ${\cal O}_{\rm LS}$ to the total decay width vanishes, for detail see e.g.\ Ref.~\cite{Dassinger:2006md}.}. Namely
\begin{align}
2 m_B \, \mu_\pi^2 (B_q) 
& =  
- \langle B_q |\bar{b}_v (i D_\mu)(i D^\mu) b_v | B_q \rangle  \, ,
\label{eq:Okin}
\\
2 m_B \, \mu_G^2 (B_q) 
& =  
\langle B_q | \bar{b}_v (i D_\mu)(i D_\nu) (-i \sigma^{\mu \nu}) b_v | B_q \rangle 
\, , 
\label{eq:Omag}
\\
2 m_B \, \rho_D^3 (B_q) & =  
\langle B_q | \bar{b}_v (i D_\mu)(i v \cdot D) (i D^\mu) b_v | B_q \rangle
\, .
\label{eq:OD}
\end{align}
Note that following Ref.~\cite{Dassinger:2006md}, the operators in Eqs.~\eqref{eq:Okin} - \eqref{eq:OD} are defined in terms of $b_v(x)$ and not the HQET field $h_v(x)$. Nevertheless, as we have explicitly checked, any differences between these two choices arise only at order $1/m_b^4$.
The values of the non-perturbative parameters $\mu_\pi^2$, $\mu_G^2$, and $\rho_D^3$ for the case of the $B^{+,0}$ meson can be determined from fits to the experimental data on
inclusive semileptonic $B \to X_c \, \ell \bar \nu_\ell$ decays.
Analyses using moments of the lepton energy and of the invariant hadronic mass distributions 
were carried out in Refs.~\cite{Alberti:2014yda, Gambino:2016jkc},
and recently in Ref.~\cite{Bordone:2021oof}, where also the
new N$^3$LO-QCD results for the parton level decay~\cite{Fael:2020tow,Czakon:2021ybq} were included.
Very recently  a new fit \cite{Bernlochner:2022ucr} has been performed using data on the moments of the dilepton invariant mass distribution reported 
by the Belle and Belle-II collaborations \cite{Belle:2021idw,Belle-II:2022fug}.
Both analyses \cite{Bordone:2021oof, Bernlochner:2022ucr} 
yield similar results for $V_{cb}$, the kinetic mass 
$m_b^{\rm kin} (1 \GeV)$ and the parameters $\mupi$, $\muG$. 
However, there appears to be a significant difference for the value of $\rhoD$,
see Table~\ref{tab:num-input}. 
Interestingly, it turns out that the Darwin operator,
which was neglected in all previous analyses of $\tsd$,
see e.g.\ Refs.~\cite{Neubert:1996we, Kirk:2017juj}, actually yields  a large or even dominant contribution to
this ratio. 
As the origin of the discrepancy in the numerical value of
 $\rho_D^3(B)$ is not yet understood, we consider two
 scenarios for the choice of the parameters 
 $m_b^{\rm kin}$, $\mu_\pi^2(B)$, $\mu_G^2(B)$, and
 $\rho_D^3(B)$, as it is summarised  in Table~\ref{tab:num-input} of the Appendix. Specifically, we refer to scenario~A when using the input from Ref.~\cite{Bordone:2021oof}, and to scenario~B when using those of Ref.~\cite{Bernlochner:2022ucr}.\\
It is worthwhile to point out that an alternative way to determine $\rho_{D}^3$ makes use of the equations of motion (EOM) for the gluon field strength tensor. In fact, taking into account that
\begin{equation}
D^\mu G_{\mu \rho}^a = - g_s \sum_q (\bar q \gamma_\rho t^a q), \qquad [i D_\mu,[i D^\rho, i D^\mu]] = g_s D_\mu G^{\mu \rho}\,,
\end{equation}
\begin{table}[t]
\renewcommand{\arraystretch}{1.5}
\centering
\begin{tabular}{|c||c|c||c|c||c|c|}
\hline
&
\multicolumn{2}{|c||}{$\mu = 4.5$ GeV} &
\multicolumn{2}{|c||}{$\mu = 1.0$ GeV} &
\multicolumn{2}{|c|}{$\alpha_s = 1$} 
\\
\hline
$\rho_D^3 [{\rm GeV^3}]$
& VIA & \mbox{ HQET} 
& VIA & \mbox{ HQET}
& VIA & \mbox{ HQET}
\\
\hline \hline
$B^+, B_d$ &  
0.029  & 0.028 &  
0.061  & 0.059 & 
0.133  & 0.128
\\
\hline
$B_s$ &  
0.044  & 0.043 &  
0.092  & 0.090 & 
0.199  & 0.195
\\
\hline
\end{tabular}
\caption{Comparison of the values of $\rho_D^3(B_q)$ obtained using Eqs.~\eqref{eq:Darwin-Penguin}, \eqref{eq:par-peng-op}, for different choices of the renormalisation scale, and using VIA and  HQET sum rules results for the non-perturbative input.}
\label{tab:rhoD}
\end{table}
the matrix element of the Darwin operator can be written up to $1/m_b$ corrections, as
\begin{equation}
- 4 m_B \, \rho_D^3 (B_q) =  g_s^2 \langle B_q| {\cal O}_P| B_q \rangle + 
{\cal O}\left( \frac{1}{m_b} \right)\,.
\label{eq:Darwin-Penguin}
\end{equation}
In Eq.~\eqref{eq:Darwin-Penguin}, we have introduced the penguin operator 
\begin{equation}
{\cal O}_P =  (\bar h_v \gamma_\mu t^a h_v) \sum_q (\bar q \gamma^\mu  t^a q)\,,
\label{eq:O-peng}
\end{equation} 
whose matrix element can be parametrised as \cite{King:2021jsq} \footnote{Note that by means of Fierz transformations, the matrix element of the penguin operator can be recast in terms of a liner combination of the four-quark operators given in Eqs.~\eqref{eq:O1-HQET} - \eqref{eq:T2-HQET}, together with the corresponding ones obtained by replacing
$(1-\gamma_5) \to (1+\gamma_5)$. Taking into account parity conservation in QCD and the parametrisation in Eqs.~\eqref{eq:ME-dim-6-HQET-q-q} - \eqref{eq:ME-dim-6-HQET-q-q-prime} leads to an expression analogous to Eq.~\eqref{eq:par-peng-op} and numerically consistent with it.}
\begin{equation}
\langle B_q| {\cal O}_P| B_q \rangle = - \frac29 f_{B_q}^2 m_B^2 \left( \tilde B_P^q + \sum_{q^\prime \neq q} \tilde \delta_P^{q^\prime q}\right) \,.
\label{eq:par-peng-op}
\end{equation}
Again, the input $\tilde B_P^q$ and $ \tilde \delta_P^{q^\prime q}$ in Eq.~\eqref{eq:par-peng-op} denote  the corresponding Bag parameters and the eye-contractions also determined for the first time in Ref.~\cite{King:2021jsq} within the framework of HQET sum rules. Note that in VIA, Eqs.~\eqref{eq:Darwin-Penguin}, \eqref{eq:par-peng-op} lead to the following simplified expression for the Darwin parameter up to power corrections, i.e.
\begin{equation}
\rho_D^3 (B_q) \approx  
\frac{g_s^2}{18} f_{B_q}^2 \, m_{B_q} \, .
\label{eq:EoM-Darwin-VIA}
\end{equation}
A comparison of the values of $\rho_D^3$ obtained for different choices of the renormalisation scale at which the strong coupling is evaluated, and using both the HQET sum rules and the VIA results for the non-perturbative input in Eq.~\eqref{eq:par-peng-op}, is shown in Table~\ref{tab:rhoD}. 
In this respect, two comments are in order. 
First, HQET sum rules and VIA yield very similar results
for $\rho_D^3$.
Second, it is interesting to observe that setting $\mu \sim m_b$,
leads to a value for $\rho_D^3$ in very good agreement
with the one obtained by the fit of Ref.~\cite{Bernlochner:2022ucr}, whereas the result given in Ref.~\cite{Bordone:2021oof} 
is reproduced by setting $\alpha_s \sim 1$, i.e.\ choosing a very low renormalisation scale.
\\
In order to predict $\Gs$ and $\tsd$,
we also need to fix the size of the largely unknown SU(3)$_F$ breaking effects in 
the non-perturbative parameters discussed above.
A possible estimate for the value of $\mu_G^2(B_{(s)})$ can be obtained using the spectroscopy relation~\cite{Uraltsev:2001ih} 
\begin{equation}
 \mu_G^2 (B_{(s)})  
 \approx \frac 3 4 \left(M_{B_{(s)}^*}^2 - M_{B_{(s)}}^2 \right) \,,
\label{eq:muG_spectr-rel}
\end{equation}
which yields
\begin{equation}
\frac{\muGs}{\muG}  
\simeq \frac{M_{B_{s}^*}^2 - M_{B_{s}}^2}{M_{B^*}^2 - M_{B}^2} 
\simeq 1.09 \pm 0.05 \,.
\label{eq:SU3f-break-muG-sr}
\end{equation}
In Eq.~\eqref{eq:SU3f-break-muG-sr}, the values of the mesons masses are taken from Ref.~\cite{Zyla:2020zbs} and lead to a vanishing uncertainty.
Therefore we have assigned a conservative $50 \%$ uncertainty 
to the deviation from one in order to account for missing power corrections.
Recently, the size of the SU(3)$_F$ breaking in $\mu_G^2$ has been estimated using lattice QCD in Refs.~\cite{Gambino:2017vkx, Gambino:2019vuo}, indicating larger effects \cite{Bordone:2022qez}
\begin{equation}
\frac{\muGs}{\muG} \simeq 1.20 \pm 0.10 \,,
\label{eq:SU3f-break-muG-Lattice}
\end{equation}
however 
in agreement with Eq.~\eqref{eq:SU3f-break-muG-sr} within uncertainties. 
In the case of the kinetic operator, $\mu_\pi^2$ can be determined using the spin-averaged masses,
see e.g. Ref.~\cite{Bigi:2011gf}, leading to \cite{Bigi:2011gf, Bordone:2022qez}
\begin{equation}
\mupis - \mupi \approx (0.04 \pm 0.02) \GeV^2 \, ,
\label{eq:SU3f-break-mupi-sr}
\end{equation}
valid up to power corrections, which are  
"accounted" by adding again a conservative $50\%$ uncertainty in Eq.~\eqref{eq:SU3f-break-mupi-sr}.   
On the other side, recent lattice QCD estimates \cite{Gambino:2019vuo} again predict somehow larger 
SU(3)$_F$ breaking effects,
\begin{equation}
\mupis - \mupi \approx (0.11 \pm 0.03) \GeV^2 \, . 
\label{eq:SU3f-break-mupi-Lattice}
\end{equation}
Finally, the size of the SU(3)$_F$ breaking effects in $\rho_D^3$ can be estimated using the EOM relation in Eq.~\eqref{eq:EoM-Darwin-VIA}, 
yielding
\begin{equation}
\frac{\rhoDs}{\rhoD} 
\approx \frac{f_{B_s}^2 \, m_{B_s}}{f_B^2 \, m_B}
\approx 1.49 \pm 0.25, 
\label{eq:SU3f-break-rhoD-EoM}
\end{equation}
where we have used lattice results for the decay constants
\cite{Aoki:2019cca}, see Table~\ref{tab:num-input},
and have again assigned additional conservative $50\%$ uncertainty
to the deviation from one due to missing power corrections.
As one can see, Eq.~\eqref{eq:SU3f-break-rhoD-EoM}
predicts very large $\approx 50$~\% SU(3)$_F$ breaking in $\rho_D^3$.
An alternative way to estimate ${\rhoDs}/{\rhoD}$ is based on the sum rules in the heavy quark limit, see e.g. Ref.~\cite{Bigi:2011gf}, which leads to
\begin{equation}
\frac{\rhoDs}{\rhoD} \simeq 
\left(\frac{\mupis}{\mupi} \right)^{\! 2}
\frac{\bar \Lambda}{\bar \Lambda_s} 
\approx 
\left\{
\begin{array}{cc}  1.05 \pm 0.09 \,  & \mbox{ using} \, \, \mbox{Eq.~\eqref{eq:SU3f-break-mupi-sr}} \, , \\[4mm]
  1.35  \pm  0.16  & \mbox{ using Eq.~\eqref{eq:SU3f-break-mupi-Lattice}} \,  ,
\end{array}
\right.
\label{eq:SU3f-break-rhoD-SR}
\end{equation}
with $\bar{\Lambda}_{(s)} = m_{B_{(s)}} - m_b$,
and we have used for $\mupi$ the value from Ref.~\cite{Gambino:2019vuo}, as it is more precise than the one from Ref.~\cite{Bernlochner:2022ucr}, cf. Table~\ref{tab:num-input}.
Note that Eq.~\eqref{eq:SU3f-break-rhoD-SR} is very
sensitive to the SU(3)$_F$ breaking in $\mu_\pi^2$, and
using the estimate in Eq.~\eqref{eq:SU3f-break-mupi-sr}
yields a much smaller value for 
$\rhoDs/\rhoD$ than the result in Eq.~\eqref{eq:SU3f-break-rhoD-EoM}.\\
As we see from the discussion above, 
the estimates of the SU(3)$_F$ breaking in the matrix elements of the two-quark operators differ quite sizeably depending on the method used.
These differences will strongly affect the lifetime ratio $\tsd$.
 Taking into account also the two fits
\cite{Bordone:2021oof, Bernlochner:2022ucr},
in our analysis we therefore consider the following two scenarios,
to cover both extreme cases for $\tsd$:
\begin{itemize}
    \item {\bf Scenario A}: 
    Non-perturbative parameters from fit by Ref.~\cite{Bordone:2021oof} 
    (with large value of~$\rho_D^3$) and larger SU(3)$_F$ breaking effects 
    by using Eqs.~\eqref{eq:SU3f-break-muG-Lattice}, \eqref{eq:SU3f-break-mupi-Lattice} and \eqref{eq:SU3f-break-rhoD-EoM}. 
    
    \item {\bf Scenario B}:
    Non-perturbative parameters from fit by Ref.~\cite{Bernlochner:2022ucr} (with small value of~$\rho_D^3$) and smaller SU(3)$_F$ breaking effects
    by Eqs.~\eqref{eq:SU3f-break-muG-sr}, \eqref{eq:SU3f-break-mupi-sr} and the first line of \eqref{eq:SU3f-break-rhoD-SR}.
\end{itemize}
A future, more precise determination of these non-perturbative parameters 
and of the corresponding size of SU(3)$_F$ breaking - 
either by fits of inclusive semileptonic $B$ and $B_s$ decays or by non-perturbative calculations - 
is clearly necessary to obtain more insights on the theoretical prediction  of $\tsd$. All the values used in our analysis for the non-perturbative
parameters, in correspondence of these two scenarios,
are summarised in Table~\ref{tab:num-input}.
\\

\section{Numerical analysis and results}
\label{sec:Res}
In this section we present the theoretical predictions for the $B$-mesons total decay widths, together with their lifetime ratios. All the input used in our analysis are listed in Appendix~\ref{app:1}. 
Note that, we consider two scenarios A and B, defined 
explicitly in the previous section, depending on which input 
we use for the parameters $\mu_\pi^2, \mu_G^2$ and $\rho_D^3$.
To be consistent with the results of both fits by 
Refs.~\cite{Bordone:2021oof} and \cite{Bernlochner:2022ucr},
by default, in our numerical analysis the mass of the $b$-quark is expressed in the kinetic scheme \cite{Bigi:1994ga, Bigi:1996si} fixing the cut-off scale $\mu^{\rm cut} = 1\, $GeV, while we adopt the $\overline{\rm MS}$ scheme for the charm quark mass \cite{Bardeen:1978yd}, i.e.
\begin{equation}
      m_b^{\rm pole}  = m_b^{\rm kin}(\mu^{\rm cut})
      \left[ 1 + \frac{4 \alpha_s}{3 \pi} 
      \left( 
      \frac 4 3 \frac{\mu^{\rm cut}}{ m_b^{\rm kin} }
      +
      \frac 1 2 \left(\frac{\mu^{\rm cut}}{ m_b^{\rm kin}}\right)^{\!\!2}
      \right)\right] + {\cal O}(\alpha_s^2)\, ,
      \label{eq:Pole-Kin-scheme}
      \end{equation}
\begin{equation}
m_c^{\rm pole} = \overline{m}_c (\overline{m}_c)
\left[1 - \frac{\alpha_s
}{\pi} 
\left(\ln \left(\frac{\overline{m}_c^2 
}{\mu^2}\right)  - \frac{4}{3}  \right) \right] + {\cal O}(\alpha_s^2)  \,.
\end{equation}
In order to understand the size of each of the contributions to the HQE, in the following we show the partial decomposition for all the five observables considered. Note that below we set $m_b^{\rm kin}(1\, {\rm GeV}) = 4.573$ GeV, and use the notation $q = u, d$, and $\tilde \delta_i^{qq} \equiv \tilde \delta_i^{ud} = \tilde \delta_i^{du}$.
For the three total widths we obtain:
\begin{eqnarray}
\Gu & = &
\Gamma_0
\biggl[ 
\underbrace{5.97}_{\rm LO} - 
\underbrace{0.44}_{\rm \Delta NLO} 
- \, 0.14 \, \frac{\mu_{\pi}^2 (B)}{\rm GeV^2}
- 0.24 \, \frac{\mu_{G}^2 (B)}{\rm GeV^2}
- 1.35 \, \frac{\rho_{D}^3 (B)}{\rm GeV^3}
\nonumber
\\[2mm]
& &
\quad 
- \, (\underbrace{0.143}_{\rm LO} + \underbrace{0.194}_{\rm \Delta NLO}) \,  {\tilde B}_1^q 
+ (\underbrace{0.023}_{\rm LO} - \underbrace{0.014}_{\rm \Delta NLO}) \,  {\tilde B}_2^q 
+ \, (\underbrace{2.29}_{\rm LO} + \underbrace{0.40}_{\rm \Delta NLO}) \,  {\tilde B}_3^q 
\nonumber \\[3mm]
& & \quad + \, (\underbrace{0.00}_{\rm LO} - \underbrace{0.05}_{\rm \Delta NLO}) \,  {\tilde B}_4^q 
- 0.01 \,  \tilde \delta^{qq}_{1} 
+ 0.01 \,  \tilde \delta^{qq}_{2} 
- 0.74 \,  \tilde \delta^{qq}_{3} 
+ 0.78 \,  \tilde \delta^{qq}_{4} 
\nonumber
\\[1mm]
& &  
\quad
- \, 0.01 \,  \tilde \delta^{sq}_{1} 
+\,  0.01 \,  \tilde \delta^{sq}_{2} 
- 0.69 \,  \tilde \delta^{sq}_{3} 
+ 0.77 \,  \tilde \delta^{sq}_{4} 
+ \underbrace{0.03}_{\rm dim.\ 7} \,
\biggr] \,,
\label{eq:Gammau}
\end{eqnarray}
\begin{eqnarray}
\Gd & = &
\Gamma_0
\biggl[ 
\underbrace{5.97}_{\rm LO} - 
\underbrace{0.44}_{\rm \Delta  NLO} 
- \, 0.14 \, \frac{\mu_{\pi}^2 (B)}{\rm GeV^2}
- 0.24 \, \frac{\mu_{G}^2 (B)}{\rm GeV^2}
- 1.35 \, \frac{\rho_{D}^3 (B)}{\rm GeV^3}
\nonumber
\\[2mm]
& &
\quad 
- \, (\underbrace{0.012}_{\rm LO} + \underbrace{0.022}_{\rm \Delta NLO}) \,  {\tilde B}_1^q 
+ (\underbrace{0.012}_{\rm LO} + \underbrace{0.020}_{\rm \Delta NLO}) \,  {\tilde B}_2^q 
- \, (\underbrace{0.74}_{\rm LO} + \underbrace{0.03}_{\rm \Delta NLO}) \,  {\tilde B}_3^q 
\nonumber \\[3mm]
& & \quad + \, (\underbrace{0.78}_{\rm LO} - \underbrace{0.01}_{\rm \Delta NLO}) \,  {\tilde B}_4^q 
- 0.14 \,  \tilde \delta^{qq}_{1} 
+ 0.02 \,  \tilde \delta^{qq}_{2} 
- 2.29 \,  \tilde \delta^{qq}_{3} 
+ 0.00 \,  \tilde \delta^{qq}_{4} 
\nonumber
\\[1mm]
& &  
\quad
- \, 0.01 \,  \tilde \delta^{sq}_{1} 
+ 0.01 \,  \tilde \delta^{sq}_{2} 
- 0.69 \,  \tilde \delta^{sq}_{3} 
+ 0.78 \,  \tilde \delta^{sq}_{4} 
+ \underbrace{0.00}_{\rm dim.\ 7}\,
\biggr] \,,
\label{eq:Gammad}
\end{eqnarray}
\begin{eqnarray}
\Gs & = &
\Gamma_0
\biggl[ 
\underbrace{5.97}_{\rm LO} - 
\underbrace{0.44}_{\rm \Delta NLO} 
- \, 0.14 \, \frac{\mu_{\pi}^2 (B_s)}{\rm GeV^2}
- 0.24 \, \frac{\mu_{G}^2 (B_s)}{\rm GeV^2}
- 1.35 \, \frac{\rho_{D}^3 (B_s)}{\rm GeV^3}
\nonumber
\\[2mm]
& &
\quad 
- \, (\underbrace{0.016}_{\rm LO} + \underbrace{0.034}_{\rm \Delta NLO}) \,  {\tilde B}_1^s 
+ (\underbrace{0.018}_{\rm LO} + \underbrace{0.033}_{\rm \Delta NLO}) \,  {\tilde B}_2^s
- \, (\underbrace{1.03}_{\rm LO} - \underbrace{0.03}_{\rm \Delta NLO}) \,  {\tilde B}_3^s 
\nonumber \\[2mm]
& & \quad + \, (\underbrace{1.16}_{\rm LO} - \underbrace{0.07}_{\rm \Delta NLO}) \,  {\tilde B}_4^s 
-  0.23 \,  \tilde \delta^{qs}_{1} 
+ 0.05 \,  \tilde \delta^{qs}_{2} 
+ 2.32 \,  \tilde \delta^{qs}_{3} 
+ 1.17 \,  \tilde \delta^{qs}_{4} 
+ \underbrace{0.00}_{\rm dim.\ 7 } \,
\biggr] \,,
\label{eq:Gammas}
\end{eqnarray}
where the size of the NLO-QCD corrections to the partonic level decay and to the coefficients of the dimension-six four-quark operators has been explicitly indicated. As we can see, the dominant contribution to Eqs.~\eqref{eq:Gammau} - \eqref{eq:Gammas} is given 
by the total width of the $b$-quark, while power and radiative corrections appear to be under control and are of order of few percents. 
 For the leading term $\Gamma_b$, as it has already been mentioned above, even N$^3$LO-QCD corrections to semileptonic decays
are known; in the scheme $m_b^{\rm kin}(1\GeV)$ and $\overline{m}_c (2 \GeV)$, the authors of Ref.~\cite{Fael:2020tow} find respectively
 $- 8.7$\% at NLO, $-1.8$\% at NNLO and $-0.03$\% at N$^3$LO, 
 meaning that higher order effects are negative and very small.
NLO-QCD corrections to $\Gamma_b$ in the total width amount to 
$\sim - 7.4 \%$, but for the actual size of higher order effects 
the complete computation of $\alpha_s^2$ corrections due to 
non-leptonic $b$-quark decays is needed.
Looking at the effect of the two-quark operators, 
it might come as surprising that the coefficient of $\rho_D^3$ is about one order of magnitude larger that those in front of $\mu_\pi^2$ and $\mu_G^2$. However, this follows from an accidental suppression of the dimension-five contribution, more than from an enhancement of the Darwin term, hence we do not expect problems with the convergence of the HQE when including higher power corrections. The series, in the case of inclusive semileptonic $B$-decays, where $1/m_b^4$ corrections to two-quark operators are known, is in fact well-behaving, see e.g.\ the discussion in Ref.~\cite{Lenz:2020oce}. Moreover, we refer to the latter for details on the size of the coefficient of the Darwin operator.
Assuming the absence of unexpected enhancements at higher orders, in our study we conservatively add extra 15$\%$ uncertainty to the dimension-six contribution in order to account for missing $1/m_b^4$ corrections.
Concerning the effect of four-quark operators, we find, as expected, that the most sizeable shift derives from the PI topology, which constitutes the dominant contribution to the total width of the $B^+$ meson, whereas it enters $\Gamma(B_d)$ and $\Gamma(B_s)$ only through the eye-contractions. On the other side, the WE diagrams, which represent the dominant topologies for the $B_d$ and $B_s$ mesons, are affected by helicity suppression at LO-QCD and lead only to a small contribution.
Furthermore, we see that while the operators $\tilde O_3^q$ and $\tilde O_4^q$ have generally the largest coefficients,
the corresponding
Bag parameters are very small, not more than few percents in magnitude, and they completely vanish in VIA.
On the contrary, the operators $\tilde O_1^q$ and $\tilde O_2^q$ have smaller coefficients but larger Bag parameters of the order of one. Regarding the size of the NLO-QCD corrections to spectator effects, they turn out to be very sizeable, particularly in the 
case of $\Gamma(B^+)$, therefore the computation of higher order corrections is of great relevance and should be 
addressed in the future. On the other side, the contribution of the eye-contractions is generally found to be negligible, if one uses the HQET sum rule predictions for the corresponding matrix elements, see Ref.~\cite{King:2021jsq}. Dimension-seven corrections are also relatively small in $\Gamma(B^+)$, where they are dominated by the PI topology,
and almost negligible for the total width of the $B_{d}$ and $B_{s}$ mesons, again because of the helicity suppression in WE. Finally, along the same line as for the
two-quark operator contribution, also in this case we add extra 15$\%$ uncertainty to the dimension-seven correction in order to account for missing higher order terms.\\
The theoretical predictions for the lifetime ratios are obtained from Eq.~\eqref{eq:ratio1} using as input the experimental value for $\tau(B^+)$ and $\tau(B_s)$ in order to cancel the dependence on the dimension-three contribution. We respectively obtain
\begin{eqnarray}
\tud
      & = & 
    1 + \, 0.059 \,  {\tilde B}_1^q + 0.005 \, {\tilde B}_2^q - 0.674 \,  {\tilde B}_3^q + 0.160 \,  {\tilde B}_4^q
   \nonumber \\[3mm] 
   & &
    - \, 0.025 \, \tilde \delta_1^{q q} 
    + 0.002 \, \tilde \delta_2^{q q} 
    + 0.591 \, \tilde \delta_3^{q q}
    - 0.152 \, \tilde \delta_4^{q q} 
   - \underbrace{0.007}_{\rm dim.\ 7}\,,
   \label{eq:tud}
\\[4mm]
\tsd
      & = & 
    1  
    + 0.026 \,\left[\mu_{\pi}^2\left(B_s\right) - \mu_{\pi }^2(B) \right] 
    + 0.043 \,\left[\mu_{G}^2\left(B_s\right) - \mu_{G }^2(B) \right]  
    \nonumber \\[3mm]
    & & 
    + \, 0.244  \, 
    \left[\rho_{D}^3\left(B_s\right) - \rho_{D}^3(B) \right] 
   \nonumber \\[2mm]
   & &
    - \, 0.0061 \, \tilde B_1^q  
    + 0.0058 \, \tilde B_2^q 
    - 0.1382 \, \tilde B_3^q
    + 0.1385 \, \tilde B_4^q
   \nonumber \\[2mm] 
   & & 
    + \, 0.0091 \, \tilde B_1^s  
    - 0.0093 \, \tilde B_2^s 
    + 0.1812 \, \tilde B_3^s 
    - 0.1967 \, \tilde B_4^s
   \nonumber \\[2mm]
   & &
    - \, 0.0258 \, \delta_1^{q q} 
    + 0.0041 \, \delta_2^{q q} 
    + 0.4136 \, \delta_3^{q q}
    + 0.0001 \, \delta_4^{q q} 
   \nonumber \\[2mm]
   & &
    - \, 0.0020 \, \tilde \delta_1^{q s} 
    +  0.0022 \, \tilde \delta_2^{q s} 
    - 0.1253 \, \tilde \delta_3^{q s}  
    + 0.1403 \, \tilde \delta_4^{q s}
   \nonumber \\[2mm]
   & &
   + \, 0.0417 \, \tilde \delta _1^{s q} 
   - 0.0095 \, \tilde \delta_2^{s q} 
   - 0.4190 \, \tilde \delta_3^{s q}  
   - 0.2110 \, \tilde \delta_4^{s q} 
   - \underbrace{7 \times 10^{-6}}_{\rm dim.\ 7}\,.
\end{eqnarray}
From Eq.~\eqref{eq:tud} and taking into account the values of the Bag parameters, we see that the dominant correction to $\tud$ is given by 
the operators ${\tilde O}_1^u$ and ${\tilde O}_3^u$
because of the large PI contribution in $\Gamma(B^+)$, 
whereas the effect of the remaining non-perturbative input is much smaller.
Therefore, in order to improve the theoretical prediction for $\tud$, both the
computation of NNLO corrections to the four-quark operators and a more precise determination
of the corresponding Bag parameters would be highly desirable.
For the ratio $\tsd$, the situation is less trivial.
The theoretical prediction for this observable is entirely
driven by the size of the SU(3)$_F$ breaking effects in the non-perturbative matrix elements
of the $B_s$ and $B_d$ mesons.
For both of them the dominant contribution
from the four-quark operators originates from the WE topology.
However, the latter is extremely suppressed, because of,
on one side, the specific combination of the
corresponding $\Delta B = 1$ Wilson coefficients, and, on
the other side, the helicity suppression affecting these
diagrams. In light of this, the role of two-quark operators becomes crucial.
Depending on the numerical input values we are using for
$\mu_\pi^2$, $\mu_G^2$ and $\rho_D^3$, the dimension-five two-quark operators can give contributions of up to several per mille to $\tsd$, while the Darwin operator can even contribute with up to two per cent,
a result absolutely unexpected a priori.

\begin{table}[t]\centering
\renewcommand{\arraystretch}{1.7}
\begin{tabular}{|C{3.0cm}|C{4.0cm}|C{4.0cm}|C{3.0cm}|}
\hline
Observable & HQE Scenario A  &  HQE Scenario B & Exp. value \\
\hline
\hline
$\Gu [{\rm ps}^{-1}]$ 
& $0.563^{+0.106}_{-0.065} $
& $0.576^{+0.107}_{-0.067} $
& $0.6105 \pm 0.0015 $
\\
\hline
$\Gd [{\rm ps}^{-1}]$ 
& $0.615^{+0.108}_{-0.069} $
& $0.627^{+0.110}_{-0.070} $
& $0.6583 \pm 0.0017 $
\\
\hline
$\Gs [{\rm ps}^{-1}]$ 
& $0.597^{+0.109}_{-0.069} $
& $0.625^{+0.110}_{-0.071} $
& $0.6596 \pm 0.0026 $
\\
\hline
\hline
$\tud $ 
& $1.0855^{+0.0232}_{-0.0219} $
& $1.0851^{+0.0230}_{-0.0217} $
& $1.076 \pm 0.004 $
\\
\hline
$\tsd $ 
& $1.0279^{+0.0113}_{-0.0113} $
& $1.0032^{+0.0063}_{-0.0063} $
& $0.998 \pm 0.005 $
\\
\hline
\end{tabular}
\caption{Theoretical predictions for the $B$-meson total decay widths and their lifetimes ratios, based on the HQE
and in correspondence of the scenarios A and B as discussed in the text. The quoted uncertainties include the variation of all the input parameters and of the renormalisation scales $\mu_1$ and $\mu_0$, together with the estimate of higher order power corrections, all combined in quadrature.
The respective experimental determinations are also shown.}
\label{tab:summary-with-uncertainties}
\end{table}
\begin{figure}[t]
    \centering
    \includegraphics[scale=1.3]{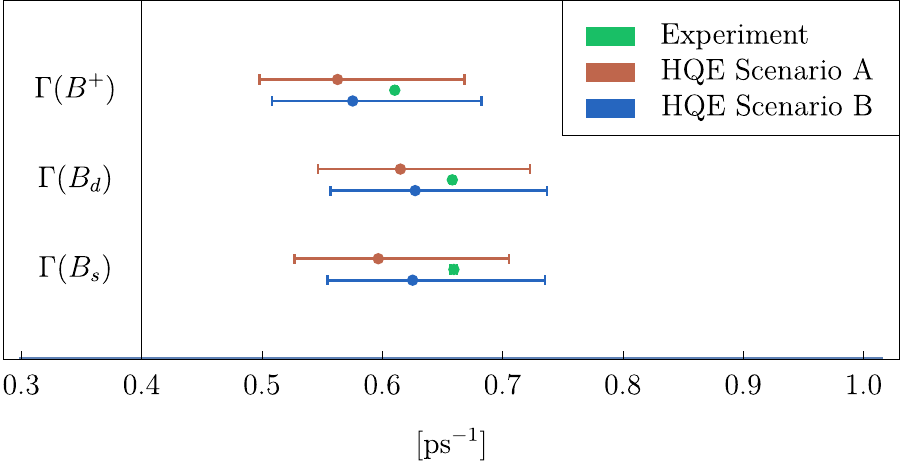}\\
    \includegraphics[scale=1.3]{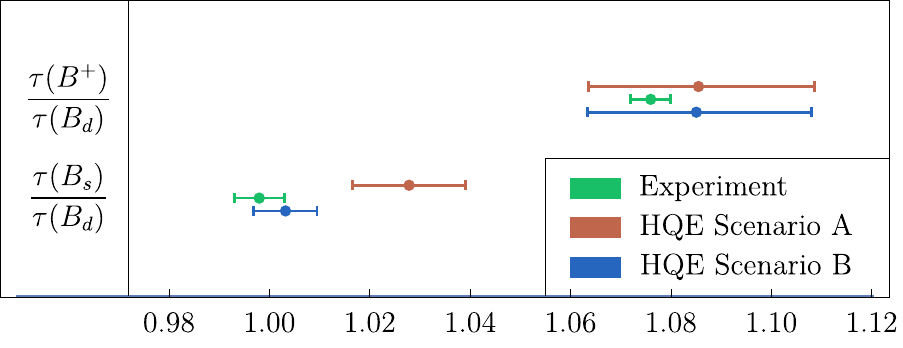}
    \caption{Summary of the HQE predictions for  $B$-meson lifetimes and their ratios within scenarios A and B defined in Section~\ref{sec:Non-pert-input}. The theoretical determinations are compared with the corresponding experimental data, 
    all the respective values are summarised in Table~\ref{tab:summary-with-uncertainties}. }
    \label{fig:results}
\end{figure}
\begin{figure}[t]
    \centering
    \includegraphics[scale=0.7]{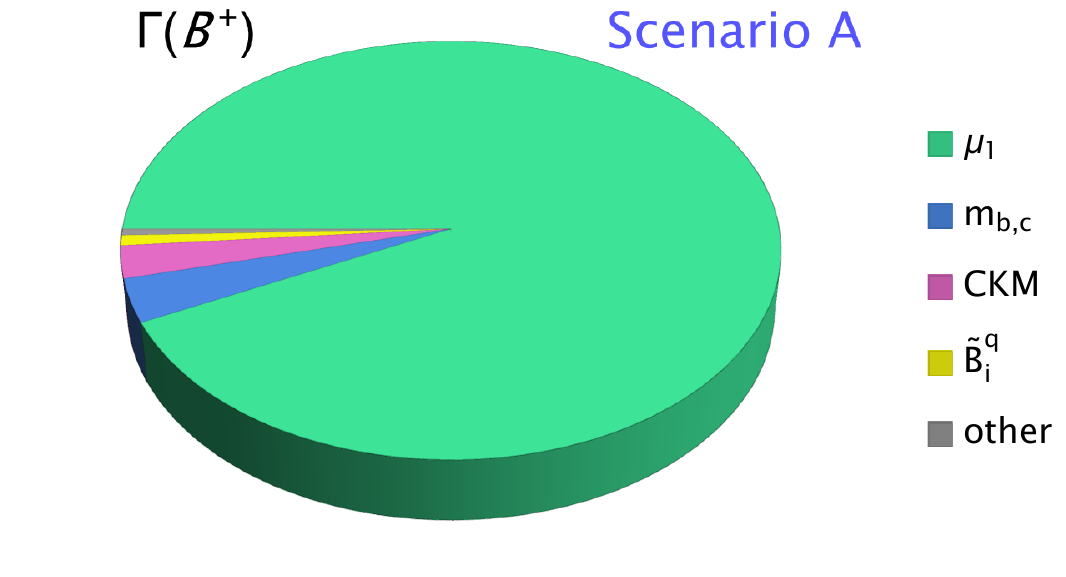}
    \includegraphics[scale=0.7]{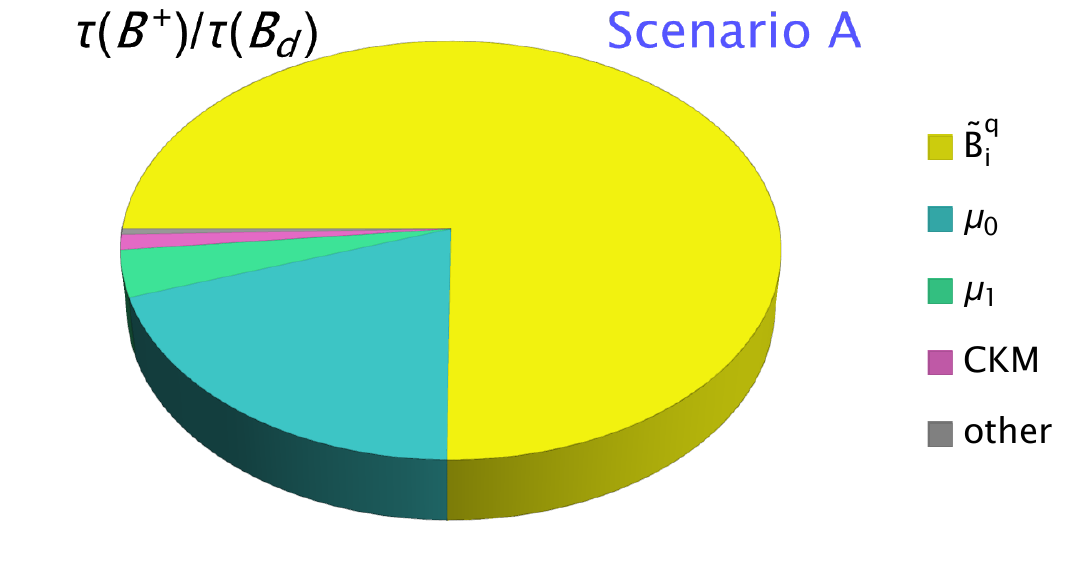}
    \\
    \includegraphics[scale=0.7]{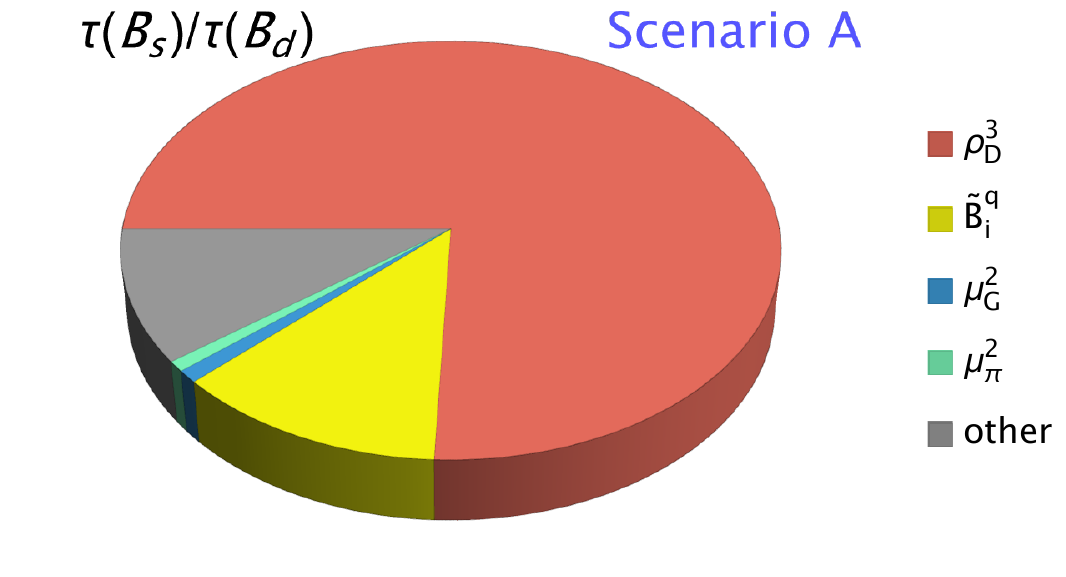}
    \includegraphics[scale=0.7]{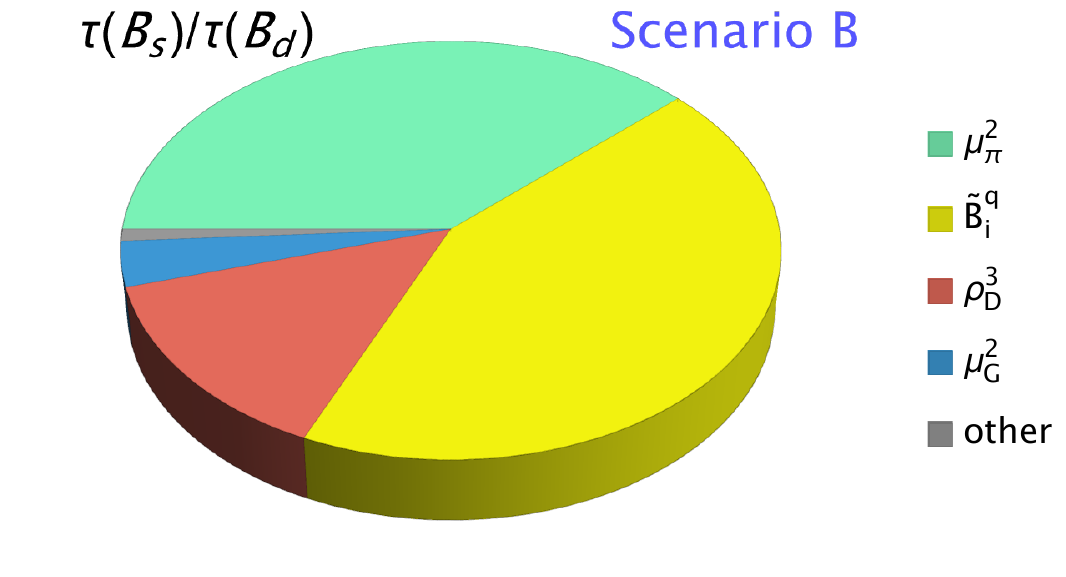}
    \caption{ Size of the individual contributions 
    to the error budget for the decay widths and lifetime ratios.
    Since we combine all uncertainties in quadrature, 
    the values shown here are defined as $(\Delta_i f/\Delta_{\rm T} f)^2$, where $\Delta_i f$ denotes an individual uncertainty due to the $i$ parameter, 
    and $\Delta_{\rm T} f$ is the total error.
    Among the decay widths, here we display only $\Gu$,
    since the pie charts for $\Gd$ and $\Gs$ are very similar
    with even smaller effects due to the Bag parameters.
    There are also no visible differences between scenarios A and B for the decay widths.
    For $\tau(B_s)/\tau(B_d)$ we show the error budget
    in both the scenarios A and B, while for $\tud$ we present only the scenario~A, since the scenario B yields almost the same picture.}
    \label{fig:pie-charts}
\end{figure}

\noindent
When varying all the input parameters within their
uncertainties, as listed in Appendix~\ref{app:1}, we
obtain the results shown in 
Table~\ref{tab:summary-with-uncertainties} and 
Fig.~\ref{fig:results}. 
 In addition, to demonstrate the size of the individual
 contribution of different parameters to the error budget for the decay widths and lifetime ratios, we show the corresponding pie charts in Fig.~\ref{fig:pie-charts}.
 We find that the main source of uncertainty to the decay widths comes from the variation of the scale $\mu_1$,  and that "subdominant" contributions 
are due to the $b$- and $c$-quark masses
and $|V_{cb}|^2$.
In the case of the lifetime ratio $\tud$ the error budget is dominated by the value of the Bag parameters and by 
the variation of the scale $\mu_0$. In this regard, as 
already stressed above, an independent computation by 
lattice QCD of the matrix elements of the dimension-six 
four-quark operators and the complete determination of 
the NNLO-QCD corrections to the corresponding 
coefficients might be very helpful in reducing the relative uncertainty. 
The lifetime ratio $\tsd$ is very sensitive to the parameter $\rhoD$,
because of its large coefficient, and to the size of 
the SU(3)$_F$ breaking effects in all the non-perturbative input.
Unfortunately the numerical values of these non-perturbative 
matrix elements are currently badly known. 
Therefore we consider two different scenarios for the parameters
$m_b^{\rm kin}(1{\rm GeV}), \mu_\pi^2, \mu_G^2$, and $\rho_D^3$
and we observe that both sets of input yield very similar results 
for all observables, except for $\tsd$. 
\\
Within uncertainties all our predictions are found to be in perfect agreement
with the corresponding experimental data, again, with the exception of $\tsd$,
where it appears to be some tension  within Scenario A. Finally, we have
explicitly checked that computing the lifetime ratios entirely within the HQE,
i.e.\ without using the experimental values for $\tau(B^+)$ and $\tau(B_s)$ as
input, leads to very similar results as the ones stated in
Table~\ref{tab:summary-with-uncertainties} and in Fig.~\ref{fig:results},
although with slightly larger uncertainties.

\section{Conclusion}
\label{sec:Conc}
With the present work, we have updated the SM prediction, within the framework of the HQE, for the decay width of the $B^+$, $B_d$,  
and $B_s$ mesons, together with the lifetime ratios 
$\tud$ and $\tsd$. 
Compared to the previous study~\cite{Kirk:2017juj}, 
we have, in addition, included the contribution of the Darwin operator, which is sizeable and crucially affects the ratio $\tsd$, corrections to the Bag parameters due to the strange quark mass, eye-contractions, and the consistent dimension-seven four-quark operator contribution in HQET.
\\
To our best knowledge, we perform the first comprehensive study of the total decay rates
of the $B_d$, $B^+$ and $B_s$ mesons. 
Our theory predictions for the latter agree well with experiment, 
albeit with large uncertainties, see Table~\ref{tab:summary-with-uncertainties}. 
From a phenomenological point of view, 
this implies that huge BSM contributions of the order of
$10 \%$ to the total decay rate can currently not be excluded by these observables. For the ratio $\tud$, our prediction is in a very good agreement with the experimental value. Compared to the experimental relative precision of 4 per mille, we find a theory uncertainty of around 2 per cent. 
In the case of $\tsd$, the situation is, however, more complicated. 
This ratio is extremely sensitive to the value of $\rhoD$ and to the size of SU(3)$_F$ breaking in the parameter $\mu_\pi^2(B)$, 
$\mu_G^2(B)$ and $\rho_D^3(B)$. Unfortunately, neither of these non-perturbative inputs are currently well known.
Specifically, we find excellent agreement of $\tsd$ with
the data  when using the fit results for $\mupi$, 
$\muG$ and $\rhoD$ from Ref.~\cite{Bernlochner:2022ucr} 
and estimates of SU(3)$_F$ breaking from spectroscopy relations. 
In that case, we also find both the experimental and theoretical precision to be of the order of 6 per mille.
However, if we use the fit results for $\mupi$, 
$\muG$ and $\rhoD$ from Ref.~\cite{Bordone:2021oof}
and the SU(3)$_F$ breaking from lattice QCD
 and from EOM relation for $\rho_D^3$,
our theory prediction for $\tsd$ lies above the
experimental value, and we find also a theoretical
uncertainty almost twice as large as the experimental one.\\
Leaving this issue with $\rho_D^3$ aside for the moment, 
we nevertheless come to the conclusion that the HQE works very well for decays of the lightest $B$-mesons, and thus provides an additional opportunity for accurate tests of 
the Standard Model and for constraining the parameter space of certain BSM models.
In order to pursue these two the goals, several possible improvements
to the theoretical prediction of $B$-meson lifetimes and their ratios would very 
desirable in the future. Specifically:
\begin{itemize}
    \item 
    Clarification of the difference in the numerical value of $\rho_D^3(B)$ obtained in the fits performed in
    Ref.~\cite{Bordone:2021oof} and Ref.~\cite{Bernlochner:2022ucr}.
    In this respect, it would probably be worthwhile to perform
    a combined fit using all sets of experimental data on semileptonic $B$-meson decays available in the literature. 
     
    \item 
    Extraction of the matrix elements of the kinetic, the chromo-magnetic and the Darwin operators for the $B_s$ meson
    from a fit to future experimental data on  inclusive semileptonic $B_s$ decays.
    This will allow for a more robust determination 
    of the size of the SU(3)$_F$ breaking effects in the corresponding non-perturbative parameters.
    In parallel, more precise non-perturbative studies of these parameters either with lattice QCD or with sum rules would be very desirable.
    
    \item
    Computation of dimension-seven two-quark operator contribution
    to non-leptonic $b$-quark decays, $\Gamma_7^{(0)}$. 
    This might in fact provide a very important correction for the theoretical determination of $\tsd$
    in the light of crucial role played by the Darwin operator, 
    as discussed in Section~\ref{sec:Res}.
    
    \item 
    Computation of $\alpha_s^2$ corrections to the non-leptonic decays of the free
    $b$-quark, $\Gamma_3^{(2)} $. In particular, this will lead to a reduction of the theory uncertainty in the total decay widths due to the renormalisation scale variation.
    A first step in that direction 
    has been done in Ref.~\cite{Czarnecki:2005vr}.
    
    \item 
    Computation of $\alpha_s^2$ corrections to the
    coefficient of the dimension-six four-quark operators, $\tilde{\Gamma}_6^{(2)} $. This may considerably improve the theoretical predictions for the lifetime ratios, since the corresponding NLO-QCD corrections
    have been found to be quite sizeable.
    
    \item 
    Computation of $\alpha_s$ corrections to the coefficient of the dimension-seven four-quark operators, $\tilde{\Gamma}_7^{(1)} $. In fact, the additional gluon contribution will lift the helicity suppression affecting the WE and WA diagrams.
    
    \item
    First lattice determination of the matrix elements of the dimension-six four-quark operators, in order to have a cross-check of the corresponding HQET sum rules results, and in particular of the small size of eye-contractions. 
    
    \item
    First non-perturbative determination of the matrix elements of 
    the dimension-seven four-quark operators. In fact, possible sizeable deviations
    from VIA in the corresponding Bag parameters may lift the helicity suppression affecting the WE and WA diagrams
    already at LO-QCD. 
    
\end{itemize}

\section*{Acknowledgements}
The authors would like to thank Martin Beneke, Gerhard Buchalla, Matteo Fael, Thomas Mannel, Daniel Moreno, Uli Nierste, Alexey Petrov, Alexei Pivovarov and Keri Vos for helpful discussions and Matthew Kirk for providing an older version of Fig.~\ref{fig:lifetime_history}. The work of
M.L.P. is financed  by the BMBF project {\it Theoretische Methoden für LHCb und Belle II}
(Förderkennzeichen 05H21PSCLA/ErUM-FSP T04).
This research has benefited from the support of the Munich Institute for Astro- Particle and BioPhysics (MIAPbP), which is funded by the Deutsche Forschungsgemeinschaft (DFG, German Research Foundation) under Germany Excellence Strategy – EXC-2094 – 390783311.

\appendix

\section{Numerical input}
\label{app:1}
We use five-loop running for $\alpha_s (\mu)$ \cite{Herren:2017osy} 
with five active flavours at the scale $\mu \sim m_b$,
and the most recent value \cite{Zyla:2020zbs}
\begin{eqnarray}
    \alpha_s (M_Z) & = & 0.1179 \pm 0.0010.
\end{eqnarray}
The masses of $B$-mesons are known very precisely \cite{Zyla:2020zbs}
\begin{equation}
m_{B^+} = 5.27934 \GeV, \quad
m_{B_d} = 5.27965 \GeV, \quad
m_{B_s} = 5.36688 \GeV \, .
\end{equation}
For the CKM matrix elements we adopt the standard parametrisation 
in terms of $\theta_{12}, \theta_{13}, \theta_{22}, \delta$
and use as input \cite{Charles:2004jd} (online update)
\begin{eqnarray}
|V_{us}| & = & 
0.22500^{+0.00024}_{-0.00021} \, , \\
\frac{|V_{ub}|}{|V_{cb}|} & = & 
0.08848^{+0.00224}_{-0.00219} \, ,
\\
|V_{cb}|& = &
0.04145^{+0.00035}_{-0.00061} \, ,
\\[1mm]
\delta & = & 
\left(65.5^{+1.3}_{-1.2}\right)^\circ\,. 
\end{eqnarray}
Regarding the quark masses, we use the kinetic scheme for the $b$-quark and the $\overline{\rm MS}$-scheme for 
the $c$-quark:
\begin{eqnarray}
m_b^{\rm kin} (\mu^{\rm cut} = 1\GeV) = (4.573 \pm 0.012) \GeV
& & \mbox{\cite{Bordone:2021oof}}\,,
\\[2mm]
m_b^{\rm kin} (\mu^{\rm cut} = 1\GeV) = (4.56 \pm 0.02) \GeV
& & \mbox{\cite{Bernlochner:2022ucr}}\,,
\\[2mm]
\overline{m_c} (\overline{m_c}) = (1.27 \pm 0.02) \GeV 
& & \mbox{\cite{Zyla:2020zbs}}\,.
\end{eqnarray} 
The charm quark mass is then run to the scale $\mu_1 = 4.5 \GeV$
using the three-loop running implemented in the RunDec package \cite{Herren:2017osy}.
Concerning the renormalisation scales $\mu_1$ and $\mu_0$, their central value is set to $\mu_1 = \mu_0 = 4.5 \GeV$
and we vary both of them independently in the interval $2.25 \GeV \leq \mu_{1, 0} \leq 9 \GeV$. The running of the Bag parameters $\tilde B_i^q$ from $\mu_0= 1.5 \GeV$ to $\mu_0 \sim m_b$ is included using the one-loop results given in Refs.~\cite{Neubert:1996we, Kirk:2017juj}.
However, we do not include the running of the eye-contractions, as they represent already
a NLO effect.
The numerical values of the Wilson coefficients, of the decay constants, and of the non-perturbative input needed to parametrise the matrix element of the two- and four-quark operators are summarised respectively in Tables~\ref{tab:WCs},
\ref{tab:num-input} and \ref{tab:Bag-parameters}.

\begin{table}[t]
\renewcommand{\arraystretch}{1.3}
\centering
   \begin{tabular}{|C{1.8cm}||C{1.8cm}|C{1.8cm}|C{1.8cm}|C{1.8cm}|C{1.8cm}|C{1.8cm}|}
   \hline 
    $\mu _1\text{[GeV]}$ & 2.5 & 4.2 & 4.5 & 4.8 & 9 \\
    \hline
 \multirow{2}{*}{$C_1 (\mu_1)$} 
  & 1.13 & 1.08 & 1.08 & 1.07 & 1.04 \\
  & (1.17) & (1.12) & (1.11) & (1.11) & (1.07) \\
 \hline
 \multirow{2}{*}{$C_2 (\mu_1)$} 
 & $-$0.27 & $-$0.19 & $-$0.18 & $-$0.17 & $-$0.11 \\
 & ($-$0.36) & ($-$0.27) & ($-$0.26) & ($-$0.25) & ($-$0.17) \\
 \hline
 \multirow{2}{*}{$C_3 (\mu_3)$} 
 & 0.02 & 0.01 & 0.01 & 0.01 & 0.01 \\
 & (0.02) & (0.01) & (0.01) & (0.01) & (0.01) \\
 \hline
 \multirow{2}{*}{$C_4 (\mu_1)$}
 & $-$0.05 & $-$0.04 & $-$0.03 & $-$0.03 & $-$0.02 \\
 & ($-$0.04) & ($-$0.03) & ($-$0.03) & ($-$0.03) & ($-$0.02) \\
 \hline
 \multirow{2}{*}{$C_5 (\mu_1)$}
 & 0.01 & 0.01 & 0.01 & 0.01 & 0.01 \\
 & (0.01) & (0.01) & (0.01) & (0.01) & (0.01) \\
 \hline
 \multirow{2}{*}{$C_6 (\mu_1)$}
 & $-$0.06 & $-$0.04 & $-$0.04 & $-$0.04 & $-$0.03 \\
 & ($-$0.05) & ($-$0.03) & ($-$0.03) & ($-$0.03) & ($-$0.02) \\
 \hline
 $C_8^{\rm eff} (\mu_1)$ 
 & ($-$0.17) & ($-$0.15) & ($-$0.15) & ($-$0.15) & ($-$0.14) \\
 \hline
    \end{tabular} 
    \caption{Values of the Wilson coefficients 
    at NLO(LO)-QCD for different choices of $\mu_1$.}
    \label{tab:WCs}
\end{table}

\begin{table}[ht]\centering
\renewcommand{\arraystretch}{1.4}
    \begin{tabular}{|c||c|c||c|c|}
    \hline 
    Parameter 
    & $B^{+,0}$ 
    & Source
    & $B_s^0$
    & Source \\
    \hline
    \hline
     $f_{B_q}$ [GeV]
    & $0.1900 \pm 0.0013$
    & LQCD \cite{Aoki:2019cca}
    & $0.2303 \pm 0.0013$ 
    & LQCD \cite{Aoki:2019cca} 
    \\
    \hline
    $\bar \Lambda_q \, [\GeV]$ &
    $0.5 \pm 0.1 $ &
    Sum Rules \cite{King:2021jsq} &
    $0.6 \pm 0.1$ &
    Sum Rules \cite{King:2021jsq}
    \\
    \hline 
    \multirow{2}{*}{$\mu_\pi^2 (B_q)$ [GeV$^2$]}
    & $0.477 \pm  0.056 $
    & Exp. fit \cite{Bordone:2021oof}
    & $0.587 \pm  0.064 $
    & Exp. fit + Eq.~\eqref{eq:SU3f-break-mupi-Lattice} 
    \\
    & $0.43 \pm  0.24$
    & Exp. fit \cite{Bernlochner:2022ucr}
    & $0.47 \pm  0.24$ 
    & Exp. fit + Eq.~\eqref{eq:SU3f-break-mupi-sr} 
    \\
    \hline
    \multirow{2}{*}{$\mu_G^2 (B_q)$ [GeV$^2$]}
    & $0.294 \pm  0.054 $
    & Exp. fit \cite{Bordone:2021oof} 
    & $0.353 \pm  0.071 $ 
    & Exp. fit + Eq.~\eqref{eq:SU3f-break-muG-Lattice}
    \\
    & $0.38 \pm  0.07 $
    & Exp. fit \cite{Bernlochner:2022ucr} 
    & $0.41 \pm  0.08 $ 
    & Exp. fit + Eq.~\eqref{eq:SU3f-break-muG-sr}
    \\
    \hline
    \multirow{2}{*}{$\rho_D^3 (B_q)$ [GeV$^3$]}
    & $0.185 \pm  0.031 $
    & Exp. fit \cite{Bordone:2021oof}
    & $0.275 \pm  0.066 $ 
    & Exp. fit +  Eq.~\eqref{eq:SU3f-break-rhoD-EoM} \\
    & $0.03 \pm  0.02 $
    & Exp. fit \cite{Bernlochner:2022ucr}
    & $0.032 \pm  0.021 $ 
    & Exp. fit + Eq.~\eqref{eq:SU3f-break-rhoD-SR} \\
    \hline
    \end{tabular}
    \caption{Numerical values of the non-perturbative parameters used in our analysis. Note that for
    $\mu_\pi^2$, $\mu_G^2$, and $\rho_D^3$,
    the first line corresponds to Scenario A, and the second one 
    to Scenario~B, both defined in Section~\ref{sec:Non-pert-input}.}
    \label{tab:num-input}
\end{table}

\begin{table}\centering
\renewcommand{\arraystretch}{1.5}
\begin{tabular}{|c||c|c|c|c|}
\hline
${\rm HQET}, \, \mu_0 = 1.5 \, {\rm GeV}$    
&  $ \tilde B_1$ 
&  $ \tilde B_2$ 
& $ \tilde \epsilon_1$ 
& $ \tilde \epsilon_2$ 
\\
\hline
\hline
    $B^{+,0}$ 
     & $\phantom{-}1.0026^{+0.0198}_{-0.0106}$ 
     & $\phantom{-}0.9982^{+0.0052}_{-0.0066}$ 
     & $-0.0165^{+0.0209}_{-0.0346}$ 
     & $-0.0004^{+0.0200}_{-0.0326}$
\\
\hline
     $B_s^0$  
     & $\phantom{-}1.0022^{+0.0185}_{-0.0099}$ 
     & $\phantom{-}0.9983^{+0.0052}_{-0.0067}$ 
     & $-0.0104^{+0.0202}_{-0.0330}$ 
     & $\phantom{-}0.0001^{+0.0199}_{-0.0324}$
\\
\hline
\end{tabular}
\begin{tabular}{|c||c|c|c|c|}
\hline
${\rm HQET}, \, \mu_0 = 1.5 \, {\rm GeV}$    
& $ \tilde \delta_1$
& $ \tilde \delta_2$ 
& $ \tilde \delta_3$ 
& $ \tilde \delta_4$ 
\\
\hline
\hline
$\langle B_q | \tilde O^q | B_q \rangle $
& $\phantom{-}0.0026^{+0.0142}_{-0.0092}$ 
& $-0.0018^{+0.0047}_{-0.0072}$ 
& $-0.0004^{+0.0015}_{-0.0024}$ 
& $\phantom{-}0.0003^{+0.0012}_{-0.0008}$
\\
\hline
$\langle B_s |  \tilde O^q | B_s \rangle$ 
& $\phantom{-}0.0025^{+0.0144}_{-0.0093}$ 
& $-0.0018^{+0.0047}_{-0.0072}$ 
& $-0.0004^{+0.0015}_{-0.0024}$ 
& $\phantom{-}0.0003^{+0.0012}_{-0.0008}$
\\
\hline
$\langle B_q | \tilde O^s | B_q \rangle$ 
& $\phantom{-}0.0023^{+0.0140}_{-0.0091}$ 
& $-0.0017^{+0.0046}_{-0.0070}$ 
& $-0.0004^{+0.0015}_{-0.0023}$ 
& $\phantom{-}0.0003^{+0.0012}_{-0.0008}$
\\
\hline
\end{tabular}
\caption{Numerical values of the Bag parameters and 
of the eye-contractions used in our analysis 
and determined in Refs.~\cite{Kirk:2017juj, King:2021jsq}. 
Here, due to the isospin symmetry, $q$ denotes both 
either $u$ or $d$ quark.}
\label{tab:Bag-parameters}
\end{table}

\section{Dimension-seven four-quark operator contribution in VIA}
\label{app:2}
In this section we list the complete expressions for the dimension-seven contribution to the PI, WE and WA diagrams depicted in Fig.~\ref{fig:PI-WE-WA}. For non-leptonic modes, these are symmetric functions of the masses of the two internal quarks and depend on one dimensionless mass parameter $\rho = m_c^2/m_b^2$.
We stress that the matrix elements of the dimension-seven HQET operators  have been evaluated in VIA. We, respectively obtain
\begin{align}
& \Gamma_{7}^{\rm WE} (\rho, 0) 
 = 
 \Gamma_0 \,  32 \pi^2 \,
\Bigl( 3 \, C_2^2 + C_1^2 \Bigr)
\rho^2 (1 - \rho) 
\frac{f_{B_q}^2 m_{B_q} \bar \Lambda_q}{m_b^4} \, ,
\label{eq:G7-WE-rho}
\\
& \Gamma_{7}^{\rm WE} (\rho,\rho) 
 = 
\Gamma_0 \, 64 \pi^2 \, \Bigl( 3 \, C_2^2 + C_1^2 \Bigr)
\frac{\rho^2}{\sqrt{1-4\rho}}
\frac{f_{B_q}^2 m_{B_q} \bar \Lambda_q}{m_b^4} \, ,
\label{eq:G7-WE-rho-rho}
\\
& \Gamma_{7}^{\rm PI} (\rho,0) 
 = 
- \Gamma_0 \, 32 \, \pi^2 \Bigl( C_1^2 + 6\,  C_1 \, C_2 + C_2^2 \Bigr) (1 - \rho)(1 + \rho)
\frac{f_{B_q}^2 m_{B_q} \bar \Lambda_q}{m_b^4} \,,
\label{eq:G7-PI-rho}
\\
& \Gamma_{7}^{\rm PI} (\rho,\rho) 
 = 
 - \Gamma_0 \, 32 \, \pi^2 \Bigl( C_1^2 + 6\,  C_1 \, C_2 + C_2^2 \Bigr) \frac{(1 - 2 \rho - 4 \rho^2)}{\sqrt{1 - 4 \rho}}
\frac{f_{B_q}^2 m_{B_q} \bar \Lambda_q}{m_b^4} \,,
\label{eq:G7-PI-rho-rho}
\end{align}
where the corresponding results for the WA topology can be derived from those for WE by exchanging $C_1 \leftrightarrow C_2 $. Moreover, in Eqs.~\eqref{eq:G7-WE-rho} - \eqref{eq:G7-PI-rho-rho}, $\Gamma_0$ is defined in Eq.~\eqref{eq:Gamma0}, while 
\begin{equation}
   \bar{\Lambda}_q = m_{B_q} - m_b\,.
\end{equation}
Note, that for internal massless quarks, it is sufficient to 
take the limit $\rho \to 0$ in the expressions above, and that due to helicity suppression, both WE and WA results vanish in this case.
Finally, for the semileptonic modes, only the WA topology is relevant, and the corresponding expression can be obtained from Eq.~\eqref{eq:G7-WE-rho}, setting 
\begin{equation}
3 \, C_2^2 \to  1\,, \qquad  C_1 \to 0\,, \qquad 
\rho \to \eta = \frac{m_\tau^2}{m_b^2} \,.
\end{equation}

\bibliographystyle{JHEP}
\bibliography{References}

\end{document}